\documentclass[acmsmall,screen]{acmart}
\usepackage{makecell} 
\usepackage{multirow} 
\usepackage{textcomp} 
\usepackage{pifont} 
\usepackage{xspace} 
\usepackage{appendix} 
\usepackage{framed} 
\usepackage{hyperref} 
\usepackage{xcolor}

\newcommand{\myso}{\texttt{PinTrace}\xspace}

\newenvironment{summary}{
  \par\addvspace{0.3em}  
  \noindent\begin{minipage}{\linewidth}
  \begin{snugshade}
  \setlength{\leftskip}{0.3em}
  \setlength{\rightskip}{0.3em}
  \noindent
}{
  \end{snugshade}
  \end{minipage}
  \par\addvspace{0.3em}  
}

\newcounter{jsonlisting}

\definecolor{shadecolor}{gray}{0.9}

\setcopyright{acmlicensed}
\copyrightyear{2026}
\acmYear{2026}
\acmDOI{xxxx/xxxx}

\begin{document}


\title{Correct Code, Vulnerable Dependencies: A Large Scale Measurement Study of LLM-Specified Library Versions}


\author{Chengjie Wang}
\email{chengjie2021@iscas.ac.cn}
\orcid{0009-0008-4445-9709}
\authornote{Intelligent Software Research Center, Institute of Software, Chinese Academy of Sciences, Beijing, China}
\authornote{University of Chinese Academy of Sciences, Beijing, China}

\author{Jingzheng Wu}
\email{jingzheng08@iscas.ac.cn}
\orcid{0000-0001-5561-9829}
\authornotemark[1] 
\authornote{Key Laboratory of System Software (Chinese Academy of Sciences), Beijing, China}

\author{Xiang Ling}
\email{lingxiang@iscas.ac.cn}
\orcid{0000-0002-7377-7844}
\authornotemark[1] 
\authornotemark[3] 

\author{Tianyue Luo}
\email{tianyue@iscas.ac.cn}
\orcid{0000-0001-7407-8255}
\authornotemark[1] 

\author{Chen Zhao}
\email{zhaochen@iscas.ac.cn}
\orcid{0009-0005-3386-0335}
\authornotemark[1] 



\begin{abstract}
Large language models (LLMs) are now largely involved in software development workflows, and the code they generate routinely includes third-party library (TPL) imports annotated with specific version identifiers.
These version choices can carry security and compatibility risks, yet they have not been systematically studied.

We present the first large-scale measurement study of version-level risk in LLM-generated Python code, evaluating 10 LLMs on \myso, a curated benchmark of 1,000 Stack Overflow programming tasks.
LLMs tend to specify version identifiers when directly prompted at 26.83\%--95.18\%, while down to 6.45\%--59.19\% in creating a manifest file directly.
Among the specified versions, 36.70\%--55.70\% of tasks contain at least one known CVE, and 62.75\%--74.51\% of them carry Critical or High severity ratings.
In 72.27\%--91.37\% of cases, the associated CVEs were publicly disclosed before the model's knowledge cutoff.
The statistics show all models converge on the same small set of risky release versions, indicating a systemic bias rather than isolated model error.
Static compatibility rates range from 19.70\% to 63.20\%, with installation failure as the dominant cause. 
The dynamic test cases confirm the pattern by 6.49\%--48.62\% pass rates.
Further experiments confirm that these failures are attributable to version selection rather than code quality, and that externally anchored version constraints substantially reduce both vulnerability exposure and compatibility failures.
Our findings reveal LLM version selection as a first-class, previously overlooked risk surface in LLM-based development.
We disclosed these findings to the community of the evaluated models, and several confirmed the issue.
All the code and dataset have been released for open science at \url{https://github.com/dw763j/PinTrace}.
\end{abstract}



\maketitle


\section{Introduction}
\label{sec:intro}

Large language models (LLMs) have become a central fixture of modern software development.
GitHub Copilot alone surpassed 20 million users by mid-2025, with deployment across 90\% of Fortune 100 companies~\cite{github_octoverse_2025}.
Copilot now contributes an average of 46\% of all code to GitHub~\cite{github_copilot_stats}, and roughly 80\% of new developers on GitHub adopt it within their first week~\cite{github_octoverse_2025}.
AI coding assistance has moved from an optional productivity tool to an expected part of the developer workflow~\cite{10.1145/3712003,10.1145/3715754}, and with that shift comes new, largely unaudited risks in the code being produced.

The development of modern software projects depends on third-party libraries (TPLs).
Studies of large open-source ecosystems document pervasive TPL adoption, with the transitive dependency graph of a typical application spanning dozens to hundreds of packages~\cite{ponta2020detection,shen2025supply}.
In traditional development, dependency version management is a deliberate, human-controlled process in which developers consult changelogs, security advisories, and compatibility notes before committing a specific release to a manifest file~\cite{10.1145/3372297.3417232,10266387}.
When an LLM generates code, however, those same version choices are made implicitly and at scale, folded into the generated output without the developer explicitly deciding what version to use.
A pinned version may carry known Common Vulnerabilities and Exposures (CVE), silently exposing every project that adopts the generated code to catalogued security risks~\cite{liu2022demystifying}.
Equally, a pinned version may be incompatible with the generated code's API usage, rendering the snippet non-executable despite appearing syntactically correct~\cite{10.1109/TSE.2023.3278129,10.1145/3691620.3695595}.
Both failure modes are invisible to a developer who adopts LLM-generated code without independent dependency auditing.

Prior work has examined two related but orthogonal risks in LLM-generated code, yet neither addresses the version dimension.
One line of research measures whether the \emph{logic} of generated code contains security weaknesses, finding that LLMs produce snippets with Common Weakness Enumeration (CWE)-classified vulnerabilities at rates estimated between 12\% and 65\% across models and task configurations~\cite{fu2025security,dai2025comprehensive,sajadi2025security,zhao2025cwe,gao2025survey}.
A second line documents that LLMs hallucinate non-existent package names at non-trivial rates, creating a supply-chain attack surface through which malicious actors can register the hallucinated names~\cite{spracklen2025package,latendresse2024chatgpt,latendresse2025robust,ladisa2023sok}.
Neither line of work addresses the \emph{version} dimension: whether the specific release of an existing, correctly named library that an LLM recommends is safe to use and compatible with the generated code.
This gap matters because version-level risks are structural.
If LLMs converge on a small set of popular but vulnerable versions, then the same CVE propagates silently and uniformly into every project that adopts LLM-generated code referencing that library~\cite{liu2022demystifying}, regardless of what task the code was written for.

We present the first large-scale measurement study of LLM version selection behavior.
We focus on the Python ecosystem because LLM-assisted Python development is pervasive~\cite{so_survey_2025} and PyPI's rich vulnerability and release metadata makes version-level risk measurable~\cite{pypi}.
We evaluate 10 LLMs on \myso, a curated benchmark of real-world Stack Overflow programming tasks, examining three dimensions: version specification behavior, security vulnerability exposure, and static and dynamic compatibility.

Our analysis of LLM-generated code on \myso reveals systematic security and compatibility risks consistent across all ten evaluated models.
First, version-annotation behavior is governed by format affordance rather than consistent engineering intent.
Under inline-comment prompting (\emph{inline} mode), LLMs specify versions for 26.83\%--95.18\% of library references.
Under manifest-based prompting (\emph{explicit} mode), the same models specify only 6.45\%--59.19\% of references.
Across both modes, version choices concentrate on a narrow band of popular releases that lags each model's knowledge cutoff by 9 to 31 months.
Second, the security consequences are severe.
Among tasks where LLMs specify versions, 36.70\%--55.70\% contain at least one version carrying a known CVE, with 62.75\%--74.51\% of those vulnerable versions rated \textit{Critical} or \textit{High} severity.
In 72\%--91\% of cases, the associated CVEs were publicly disclosed before each model's knowledge cutoff.
Yet the same risky versions appear consistently across all model families, pointing to a shared systemic bias rather than an isolated model defect.
Third, the specified versions are frequently incompatible with the generated code.
Static compatibility rates fall to 19.70\%--63.20\% under inline prompting.
Dynamic verification confirms the pattern, with pass rates collapsing to 6.49\%--48.62\% as version incompatibilities block execution before any test can run.
Together, these three findings establish that LLM version selection introduces risks along both the security and compatibility dimensions.

We conducted three additional analyses to characterize the boundaries and root causes of these risks.
We first examine whether the compatibility findings hold across different Python runtime environments.
Compatibility rates peak at mid-range Python versions and collapse sharply at more recent releases, with representative models dropping by more than 50 percentage points across that transition.
Under explicit mode the trend reverses, with compatibility improving monotonically toward newer Python versions.
This reversal confirms that LLMs systematically prefer older library releases that are compatible with legacy environments but fail under modern runtime configurations.
A controlled diagnosis experiment then establishes that installation failures are attributable to version selection rather than to defects in the generated code itself.
A mitigation probe shows that natural-language safety instructions yield no meaningful improvement, while external version anchoring substantially reduces vulnerability exposure.
RAG-augmented grounding provides marginal additional compatibility gains, though a residual gap remains.
These analyses confirm that version-selection risk is addressable only through external tooling, not through model-level prompting.
We disclosed these findings to the community of all evaluated models and to major coding assistant providers, several of whom have acknowledged the problem.

The contributions of this paper are as follows:
\begin{itemize}
  \item A measurement study of LLM version selection behavior across ten models on our curated \myso dataset with 1,000 real-world Stack Overflow questions, covering version specification rates, validity, vulnerability exposure, and compatibility.
  \item Empirical evidence that LLM-specified versions are systematically vulnerable, converge across model families on the same small set of risky releases, and lag behind each model's knowledge cutoff due to a systemic training-signal bias.
  \item A diagnosis experiment establishing that installation-level compatibility failures are attributable to version selection rather than generated code quality, and a mitigation probe characterizing the effectiveness and limits of prompt-level interventions.
  \item A dataset and evaluation pipeline for reproducible version-level security and compatibility assessment of LLM-generated code, released at \url{https://github.com/dw763j/PinTrace}.
\end{itemize}

The remainder of the paper is organized as follows.
§\ref{sec:background} provides background on LLM-assisted development and software supply chain security.
§\ref{sec:study_design} describes the study design.
§\ref{sec:exp_setup} details the experimental setup.
§\ref{sec:results} presents the main results for RQ1--RQ3.
§\ref{sec:rdm} reports the robustness, diagnosis, and mitigation analyses.
§\ref{sec:discussion} discusses implications for developers and LLM providers.
§\ref{sec:threats} addresses threats to validity.
§\ref{sec:related} surveys related work.
§\ref{sec:conclusion} concludes.
 

\section{Background}
\label{sec:background}

This section provides the prior knowledge needed to understand our study: how LLMs have changed the dependency annotation process, and why version-level choices represent an underexplored point of failure in software supply chain security.

\subsection{LLM-Assisted Development and Dependency Annotation}

LLMs have fundamentally changed how software is written~\cite{10.1145/3695988}.
Tools such as GitHub Copilot, Cursor, and Claude Code are now embedded into the daily workflow of millions of developers~\cite{zhang2026survey}.
The 2025 Stack Overflow Developer Survey reports that 84\% of developers use or plan to use AI coding tools, and 51\% of professionals do so daily~\cite{so_survey_2025}.
These tools do not merely suggest individual lines.
They generate complete, self-contained code snippets that include import statements, function bodies, and dependency annotations in a single pass~\cite{github_octoverse_2025,li-etal-2025-fea,10.1145/3747588}.
As a result, version choices for TPLs now flow into codebases through an automated, largely unreviewed channel, at a scale and speed that traditional human-driven dependency management was not designed to handle.

\begin{figure}[t]
  \centering
  \includegraphics[width=\linewidth]{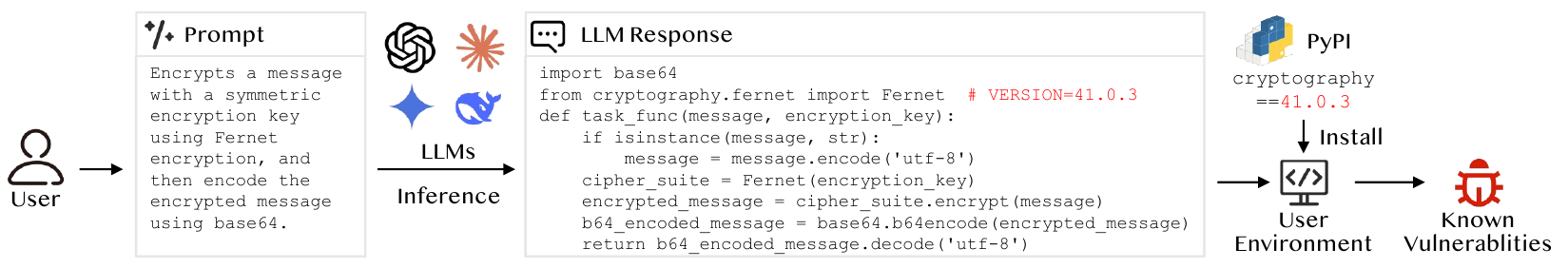}
  \Description{A motivating example showing LLM-generated Python code that imports the cryptography library and annotates it with version 41.0.3. The version installs successfully but carries 7 known CVEs, illustrating that syntactically correct, installable code can silently expose a project to catalogued vulnerabilities through its pinned dependency version.}
  \caption{A motivating example of LLM-introduced version-level risk.
           An LLM generates functionally correct code and annotates the import
           with \texttt{cryptography==41.0.3}.
           The version installs successfully, yet carries 7 known CVEs,
           silently exposing the user's environment to catalogued vulnerabilities.}
  \label{fig:motivation}
\end{figure}

When developers write code manually, selecting a library version is a deliberate act where they consult changelogs, security advisories, and compatibility notes before committing a specific release to a manifest file~\cite{10.1145/3372297.3417232,10266387}.
LLMs displace this decision with parametric inference.
The model draws on version information encoded in its training corpus, without access to live vulnerability databases or the ability to verify whether the chosen release is current, safe, or installable~\cite{latendresse2024chatgpt,latendresse2025librarian}.
A developer who accepts a generated snippet inherits the model's version choice wholesale, often without realizing a choice was made at all.
The scale of this risk has begun to surface in industry data: analysis of 36,870 LLM-assisted dependency upgrade recommendations found that 27.76\% of version suggestions were hallucinations~\cite{sonatype2026}.
Some recommendations pointed to packages that were outright malware.
Figure~\ref{fig:motivation} illustrates a concrete instance of this problem from the BigCodeBench dataset~\cite{bigcodebench}.
Given a task that requires the \texttt{cryptography} library, an LLM produces syntactically correct, functionally plausible code and annotates the import with \texttt{\# VERSION=41.0.3}.
The version exists on PyPI and installs without error, so neither the developer nor a build system raises an alert.
Yet \texttt{cryptography==41.0.3} carries 7 known CVEs, exposing the user's environment to catalogued vulnerabilities without any visible warning.
This scenario, invisible in the absence of explicit security auditing, is the central concern of this paper.

\subsection{Software Supply Chain Security and the Version Control Point}

Third-party libraries substantially improve development efficiency in the form of software supply chain~\cite{synkshift,ntiaframing}, but they also expand the attack surface of every project that uses them~\cite{shen2025supply,ladisa2023sok,10.1145/3708531,10.1145/3640336,shen2025understanding}.
At the scale of modern registries, vulnerability exposure is not an edge case: in 2025, one in five PyPI releases was associated with a vulnerability rated CVSS 7.0 or higher~\cite{sonatype2026}.
High-profile incidents, including Log4Shell~\cite{log4j}, SolarWinds~\cite{alkhadra2021solar}, and the XZ Utils backdoor~\cite{przymus2025wolves}, demonstrate that a single compromised or vulnerable version in a widely-used library can trigger systemic compromise at a global scale.
Once a vulnerability is present in a widely-used library version, every project depending on that version inherits the same exposure through the dependency graph, with no per-project variation~\cite{liu2022demystifying,10.1109/ICSE48619.2023.00095,10.1109/ICSE48619.2023.00031}.

To manage these risks, the software engineering community has developed software composition analysis (SCA) tools and software bill of materials (SBOM) frameworks that identify and track third-party components and their known vulnerabilities~\cite{ponta2020detection,imtiaz2021comparative,20252604,10.1145/3788692,hu2024empirical}.
These approaches are effective at auditing committed dependency files and flagging vulnerable versions after the fact.
However, SCA and SBOM operate post hoc.
By the time the tools run, the version has already been written into the manifest, and the window for intervention has closed.
When that manifest is generated by an LLM, the version choice is made at generation time, before any developer review or security tooling is involved.
This paper addresses that upstream gap: not whether a vulnerable version is eventually detected, but how reliably LLMs introduce one in the first place.


\section{Study Design}
\label{sec:study_design}

Though LLMs introduce version choices into codebases implicitly, those choices are not audited at the point of generation.
To characterize this risk empirically, we design a measurement study organized around three research questions.
§\ref{sec:rqs} defines the questions and the rationale connecting them.
§\ref{sec:dataset_construction} describes how we construct the task dataset.
§\ref{sec:prompting_modes} introduces the two prompting modes.
§\ref{sec:pipeline} presents the analysis pipeline.
§\ref{sec:metrics} formally defines all evaluation metrics.

\subsection{Research Questions}
\label{sec:rqs}

We organize the study around three research questions that together characterize version-level risk from its origin to its consequences.
RQ1 establishes the empirical baseline: what versions LLMs actually produce and whether those identifiers are valid.
RQ2 probes the security consequences of those choices.
RQ3 asks whether the chosen versions are executable in practice.

\noindent\textbf{RQ1. Version specification behavior:}
\textit{When generating code, do LLMs annotate TPL dependencies with version identifiers, and are the specified versions valid published releases?}
Unpinned dependencies introduce non-reproducible builds and silent upgrade risks irrespective of any known vulnerability~\cite{10.1145/3715728,10.1007/s10664-017-9521-5}.
Invalid versions cannot be installed and reveal the model's tendency to hallucinate plausible but nonexistent identifiers.
RQ1 therefore provides the empirical baseline on which RQ2 and RQ3 are built.

\noindent\textbf{RQ2. Version vulnerability exposure:}
\textit{Among the valid versions that LLMs specify, what proportion carry known CVE vulnerabilities and how are they distributed across severity levels?}
Prior work has measured whether generated code contains insecure logic~\cite{fu2025security,dai2025comprehensive,sajadi2025security,zhao2025cwe} or whether recommended packages exist~\cite{spracklen2025package,latendresse2024chatgpt,latendresse2025robust,ladisa2023sok}.
Our focus is orthogonal, asking whether the specific \emph{versions} of real, correctly named libraries expose users to catalogued vulnerabilities.
A high exposure rate signals a systemic risk, because the same CVE propagates into every project that adopts LLM-generated code referencing that library.

\noindent\textbf{RQ3. Version-Code compatibility:}
\textit{Are the TPL versions that LLMs specify compatible with the generated code?}
Even a valid, vulnerability-free version may be incompatible with the API usage in the generated code, making the snippet non-executable~\cite{10121651,11329192,10.1145/3728875}.
We decompose compatibility into two dimensions: \emph{static compatibility}, assessed via installation success and static type checking, and \emph{dynamic compatibility}, assessed via execution of test suites against the TPL versions.

\subsection{Dataset Construction}
\label{sec:dataset_construction}

To ground the study in authentic development scenarios, we build a task dataset from real-world programming Q\&A rather than synthetic benchmarks.

\textbf{Programming language.}
Rigorous vulnerability analysis and compatibility verification require a controlled execution environment, a comprehensive vulnerability database, and a reproducible package ecosystem. These constraints are best satisfied by committing to a single programming language.
We focus on Python for three reasons.
First, Python is the dominant language in data science, machine learning, and scripting, the domains most frequently targeted by LLM-assisted coding~\cite{so_survey_2025,github_octoverse_2025}.
LLM dependency recommendations therefore carry the greatest practical consequence in Python projects.
Second, PyPI hosts over 600,000 packages with comprehensive release histories and vulnerability records~\cite{sonatype2026}.
This breadth provides the infrastructure needed for version validation and vulnerability lookup at scale.
Third, PyPI's rapid release cycles, reflected in a 50.64\% year-over-year growth in downloads in 2025~\cite{sonatype2026}, produce version-level compatibility failures at a frequency that makes Python a high-signal environment for the phenomena we study.

\textbf{Data source.}
We derive tasks from Stack Overflow, a  website of Q\&A about programming.
Stack Overflow questions reflect genuine development scenarios encountered by real practitioners, and their accepted answers represent community-validated solutions that provide a meaningful reference for our analysis.

\textbf{Selection criteria.}
We filter the dump by five criteria:
\begin{itemize}
  \item The question has an accepted answer.
  \item The accepted answer contains at least one fenced code block.
  \item The code block parses successfully as a Python abstract syntax tree (AST), excluding pseudocode, shell scripts, and malformed snippets.
  \item The parsed code contains at least one \texttt{import} statement referencing a third-party library.
  \item The question was posted between January 1, 2020 and January 1, 2026.
\end{itemize}
The temporal window spans three distinct knowledge regimes.
Questions from 2020--2022, predating widespread LLM adoption, are likely to have entered LLM training corpora, providing tasks on which models may recall specific version information.
Questions from 2022--2024 coincide with active LLM deployment.
Questions from 2025 are likely beyond the knowledge cutoff of most evaluated models, probing version knowledge under distribution shift.
This spread allows us to examine whether version-annotation accuracy varies across these regimes.

\textbf{Balanced sampling.}
The TPLs referenced in Stack Overflow question-answer pairs are not uniformly distributed: commonly used libraries such as \texttt{pandas} and \texttt{numpy} may dominate the raw question volume, and naive random sampling would over-represent this small set of high-volume libraries, reducing the ecological diversity of the study.
We therefore apply TPL-balanced sampling. We restrict to posts with non-empty target libraries, let 
$\mathcal{K}$ denotes the number of distinct libraries in the pool, and draw 
$\mathcal{N}$ posts by assigning each library a target count of approximately $\mathcal{N/K}$ and greedily selecting disjoint posts to satisfy these targets where possible, then filling any shortfall by choosing posts that most reduce under-coverage of their mentioned libraries. The resulting subset is not proportional to usage frequency in the raw data, but is intended to give each library a comparable chance of inclusion and to improve ecological diversity.


\subsection{Prompting Modes}
\label{sec:prompting_modes}

To ensure measurement results are representative and cross-model comparisons are valid, we standardize two prompting modes that correspond to the two common ways developers request dependency-annotated code from an LLM.
Using controlled modes rather than free-form prompts relieves prompt-wording variance as a confound and makes results comparable across models.

\textbf{Explicit mode.}
The prompt instructs the LLM to generate both a code snippet and a companion \texttt{requirements.txt} file listing each dependency with a version specifier.
This reflects standard software-engineering practice and supports relational specifiers, \textit{e.g.}, \texttt{==}, \texttt{>=}, \texttt{\textasciitilde=}, where models may express version ranges rather than exact pins.

\textbf{Inline mode.}
The prompt instructs the LLM to annotate each third-party import with a version identifier in a trailing comment, following the convention \texttt{import \emph{lib}\ \ \# version==\emph{x.y.z}}.
This mode reflects the practice of sharing self-contained snippets in chat sessions, Jupyter notebooks, or single-file scripts, where no manifest is expected.

The two modes are complementary rather than redundant.
Explicit mode captures how developers request structured dependency management.
Inline mode captures the embedded version annotation that accompanies everyday code generation.
Together, they cover the full range of version-specification behavior that practitioners encounter. The full prompts are appended based on the basic structure of the prompt from the BigCodeBench dataset~\cite{bigcodebench}.
\subsection{Analysis Pipeline}
\label{sec:pipeline}

\begin{figure}[t]
  \centering
  \includegraphics[width=\linewidth]{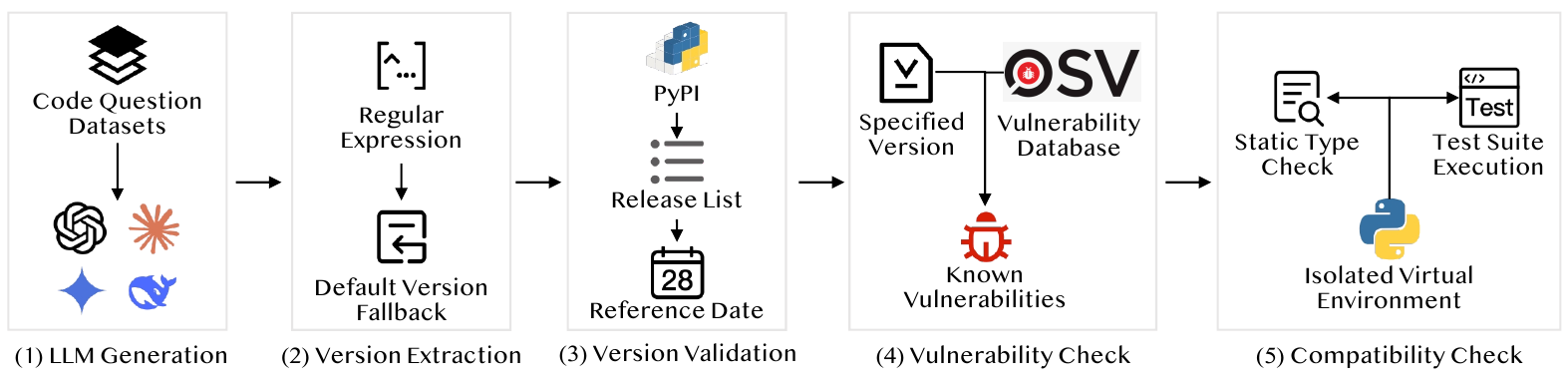}
  \Description{A flowchart illustrating the five-stage analysis pipeline of the measurement study. Each task-model-mode triple progresses through: Stage (1), code generation, where the question is submitted to the LLM; Stage (2), version extraction, where library references and version specifiers are parsed from the generated output; Stage (3), version validation, where each specified version is checked against the PyPI release history; Stage (4), vulnerability lookup, where valid versions are queried against the OSV vulnerability index to retrieve CVE records and severity scores; and Stage (5), compatibility verification, comprising both static installation and static type checking, and dynamic test suite execution on BigCodeBench. Stages (1) through (4) supply data for RQ1 and RQ2; Stage (5) supplies data for RQ3.}
  \caption{The analysis pipeline of the measurement study.
    Each (task, model, mode) triple passes through code generation, version extraction, version validation, vulnerability lookup, and compatibility verification.
    Stages (1)--(4) supply data for RQ1 and RQ2. Stage (5) supplies data for RQ3.}
  \label{fig:pipeline}
\end{figure}

We analyze version-level security and compatibility through a five-stage pipeline, illustrated in Figure~\ref{fig:pipeline}.
Each task $t \in \mathcal{T}$ is processed under every combination of model $m \in \mathcal{M}$ and prompting mode $p \in \mathcal{P} = \{\texttt{inline},\allowbreak\,\texttt{explicit}\}$, yielding the triple $(t, m, p)$ as the basic unit of analysis.
Stages (1)--(4) produce the data for RQ1 and RQ2. Stage (5) addresses RQ3.

\textbf{(1) LLM generation.}
We submit question text $q_t$ to model $m$ under prompting mode $p$ and record the raw output $G(t, m, p)$.

\textbf{(2) Version extraction.}
We parse $G(t, m, p)$ to identify all TPL references and their associated version specifiers, yielding a set of (library, version) pairs $\mathcal{V}(t, m, p) = \{(\ell_1, v_1), \ldots, (\ell_k, v_k)\}$.
Under inline mode, we apply a regular expression over import-comment patterns.
Under explicit mode, we parse the \texttt{requirements.txt} block line by line.
For libraries without any version specifier, we resolve the version to the latest PyPI release as of a unified cutoff date for all models.

\textbf{(3) Version validation.}
For each $(\ell, v) \in \mathcal{V}(t, m, p)$, we query the PyPI JSON API to verify that version $v$ is a published distribution of library $\ell$.
Versions are marked invalid if they are absent from PyPI's release history, have been yanked, or belong to packages unavailable on PyPI.

\textbf{(4) Vulnerability check.}
For each valid $(\ell, v)$, we query a pre-built vulnerability index to retrieve all CVE records affecting that specific version, along with their Common Vulnerability Scoring System (CVSS) severity classifications.
The construction of this index from the Open Source Vulnerability (OSV) dataset~\cite{osv} is described in §\ref{sec:exp_setup}.

\textbf{(5) Compatibility check.}
We assess compatibility in two sequential steps that together cover both static and dynamic dimensions.

\emph{Static type check.}
For each task $t$, we create an isolated Python virtual environment and attempt to install all TPLs in $\mathcal{V}(t, m, p)$ simultaneously.
For tasks where installation succeeds, we run a static type checker on $G(t, m, p)$ against the installed library stubs, detecting API-level incompatibilities such as removed functions or changed signatures.
A task is statically compatible if installation succeeds and the type checker reports no errors.

\emph{Dynamic execution check.}
Static type compatibility is a necessary but not sufficient condition for runtime correctness.
To obtain execution-level evidence, we integrate our pipeline with BigCodeBench~\cite{bigcodebench}, an established benchmark that pairs Python programming tasks with ground-truth test suites spanning a range of TPLs.
This integration enables static and dynamic signals to be collected on the same tasks under identical environments, providing a complementary and cross-validating view of version compatibility.
For each BigCodeBench task, we construct the same isolated environment as above and execute the provided test suite.
A task passes if all suite assertions hold.

Taken together, the pipeline transforms each $(t, m, p)$ triple into a structured record covering version specification, validity, vulnerability exposure, and compatibility, providing the empirical data for all three research questions.

\subsection{Metrics}
\label{sec:metrics}

We formally define the main metrics used in our evaluations.
Let $L(t, m, p)$ denote the multiset of all third-party library references in $G(t, m, p)$.
We define three subsets of $L$.
$L_S \subseteq L$ contains library uses that carry an explicit version specifier.
$L_V \subseteq L$ contains library uses whose resolved version corresponds to a published PyPI release, where the resolved version is either the explicitly specified version or the latest-release fallback assigned in Stage (2).
$L_U \subseteq L_V$ further restricts to versions carrying at least one associated CVE.
The filtering applied by Stages (2)--(4) determines membership in $L_S$, $L_V$, and $L_U$, respectively.

\textbf{Library-level rates} measure behavior aggregated over all library uses across tasks.
Three rates share the same form, each measuring the fraction of library uses satisfying a given criterion relative to all library uses $L$:

\begin{center}
\begin{tabular}{@{}lll@{}}
\toprule
\textbf{Metric} & \textbf{Symbol} & \textbf{Ratio} \\
\midrule
Version specification rate & $\rho_S$ & $|L_S| \mathbin{/} |L|$ \\
Version validity rate       & $\rho_V$ & $|L_V| \mathbin{/} |L|$ \\
Library vulnerability rate  & $\rho_U$ & $|L_U| \mathbin{/} |L_V|$ \\
\bottomrule
\end{tabular}
\end{center}

\noindent where all sums are taken over $t \in \mathcal{T}$ for a given $(m, p)$.

High-risk rate $(\rho_H)$ characterizes the severity profile of vulnerable versions by measuring what fraction are rated \textit{Critical} or \textit{High} by CVSS:
\begin{equation}
  \rho_H(m, p) =
  \frac{\bigl|\{(\ell,v) \in \bigcup_{t} L_U(t,m,p)
               : \mathrm{sev}(\ell,v) \in \{\textit{Critical},\textit{High}\}\}\bigr|}
       {\bigl|\bigcup_{t} L_U(t,m,p)\bigr|}
\end{equation}

\textbf{Task-level metrics} measure risk at the granularity of individual tasks rather than individual library uses.
Task vulnerability exposure $(\tau_U)$ counts the fraction of tasks in which at least one specified version carries a CVE:

\begin{equation}
\tau_U(m, p) =
  \frac{|\{t \in \mathcal{T} : L_U(t,m,p) \neq \emptyset\}|}
       {|\mathcal{T}|}
\end{equation}

This captures user-facing risk more directly than $\rho_U$, since a single vulnerable library is sufficient to expose the entire task.

Task compatibility rate $(\tau_C)$ is the fraction of tasks that pass Stage (5):
\begin{equation}
  \tau_C(m, p) =
  \frac{|\{t \in \mathcal{T} : \mathrm{compat}(t,m,p) = 1\}|}
       {|\mathcal{T}|}
\end{equation}
The same formula applies to both static and dynamic compatibility. The specific task population in each case is clarified in the corresponding results section.


\section{Experiment Setup}
\label{sec:exp_setup}

This section documents the concrete configuration of the study.
§\ref{sec:setup_dataset} reports the dataset statistics.
§\ref{sec:setup_models} lists the evaluated LLMs and their inference settings.
§\ref{sec:setup_vuln_index} describes the construction of the vulnerability index.
§\ref{sec:setup_env} specifies the execution environment for compatibility verification.

\subsection{Dataset Statistics}
\label{sec:setup_dataset}

\begin{table}[t]\small
\centering
\caption{Temporal distribution of \myso dataset.
  The three knowledge regimes reflect each year's likely relationship to LLM training corpora: pre-adoption questions are likely memorized, deployment-era and near-cutoff questions may be partially covered, post-cutoff questions probe out-of-distribution version knowledge.}
\label{tab:dataset_temporal}
\begin{tabular}{lrrrl}
\toprule
\textbf{Year} & \textbf{\# Tasks} & \textbf{Cumulating\ (\%)} & \textbf{\# Unique Libs} & \textbf{Knowledge Regime} \\
\midrule
2020 & 313 & 31.3 & 177 & Pre-LLM adoption \\
2021 & 246 & 55.9 & 153 & Pre-LLM adoption \\
2022 & 194 & 75.3 & 133 & Active LLM deployment \\
2023 & 131 & 88.4 & 101 & Active LLM deployment \\
2024 &  76 & 96.0 &  74 & Near-cutoff \\
2025 &  40 & 100.0 &  49 & Post-cutoff (most models) \\
\midrule
Total & 1,000 & 100.0 & 267 & --- \\
\bottomrule
\end{tabular}
\end{table}

\myso is constructed from the official Stack Overflow data dump~\cite{so_dump}, which archives all posts and answers on the website, and it is updated on January 6, 2026.
The full dump contains 24,178,621 questions, of which 12,392,896 carry an accepted answer.

Before applying the selection criteria, we first establish the library boundary for the study.
We restrict the candidate pool to libraries that appear in the OSV PyPI vulnerability database~\cite{osv}, \textit{i.e.}, libraries for which at least one CVE record exists in the index.
This boundary ensures that every task in the dataset is capable of producing a meaningful vulnerability signal in Stage (4), and it aligns the library scope of the dataset with the scope of the vulnerability index, eliminating a systematic blind spot in which the dataset contains libraries that the index cannot evaluate.

Within this boundary, applying the selection criteria in §\ref{sec:dataset_construction} yields 405,707 candidate records after filtering for Python context and third-party library involvement.
TPL-balanced sampling then reduces this pool to a final dataset of \textbf{1,000 tasks} covering \textbf{267 distinct third-party libraries}.

Python's import system allows a package to be imported under a name that differs from its registered name on PyPI.
For example, the \texttt{PIL} import corresponds to the \texttt{pillow} package on PyPI, and \texttt{sklearn} corresponds to \texttt{scikit-learn}.
To address this, we apply the widely adopted import-to-package name mapping included in the \texttt{pipreqs} project~\cite{pipreqs}, which covers 1,157 widely used Python packages exhibiting such discrepancies and is actively maintained by the community.

Table~\ref{tab:dataset_temporal} reports the temporal and library distribution of tasks.
The dataset is front-weighted toward 2020--2022, reflecting the higher volume of Stack Overflow activity in that period.
Tasks from 2024 onward are fewer, partly because later questions are more likely to fall beyond the knowledge cutoff of the evaluated models, and partly because the rise of LLM-based coding assistants has reduced the volume of questions posted to Stack Overflow from over 182,948 questions per month in 2020 to a maximum of 18,239 a month in 2025, which is roughly 10\% remained~\cite{so_query}.
The dataset nonetheless retains 116 tasks from 2024--2025 to probe version knowledge under distribution shift.
Unique library counts per year are broadly proportional to task counts, indicating that the temporal front-weighting does not artificially concentrate the library coverage in any single period.

The 267 libraries span a wide range of application domains, including web frameworks (\textit{e.g.}, \texttt{django}, \texttt{flask}, \texttt{fastapi}), data science and machine learning (\textit{e.g.}, \texttt{pandas}, \texttt{numpy}, \texttt{torch}), security and cryptography (\textit{e.g.}, \texttt{cryptography}, \texttt{pycryptodome}), and network utilities (\textit{e.g.}, \texttt{requests}, \texttt{paramiko}, \texttt{scapy}).
The five most frequent libraries are \texttt{pandas} (142 tasks), \texttt{numpy} (123), \texttt{django} (88), \texttt{flask} (61), and \texttt{requests} (56). While the remaining 262 libraries each appear in at most 26 tasks, confirming that balanced sampling prevents any single library from dominating the dataset.

\begin{table}[t]\small
\centering
\caption{Content statistics of the \myso dataset.
  Token counts are calculated by the GPT-5 tokenizer from the ChatGPT official tool tiktoken~\cite{tiktoken}.
  TPL import count is measured on the accepted answer's code block. The standard deviation (Std.) is computed across all 1,000 tasks.}
\label{tab:dataset_content}
\begin{tabular}{lrrr}
\toprule
\textbf{Statistic} & \textbf{Mean} & \textbf{Median} & \textbf{Std.} \\
\midrule
\multicolumn{4}{l}{\textit{Question}} \\
\ \ Token length             & 430.6 & 286.5 & 518.3 \\
\midrule
\multicolumn{4}{l}{\textit{Accepted answer code block}} \\
\ \ Lines of code            & 24.9 & 16.0 & 39.3 \\
\ \ Token length             & 202.0 & 123.5 & 298.8 \\
\ \ TPL imports per task     & 1.6 & 1.0 & 0.8 \\
\midrule
\multicolumn{4}{l}{\textit{Library coverage}} \\
\ \ Distinct TPLs            & \multicolumn{3}{r}{267} \\
\ \ Tasks with $\geq$2 TPL imports & \multicolumn{3}{r}{464 (46.4\%)} \\
\ \ Tasks with $\geq$3 TPL imports & \multicolumn{3}{r}{131 (13.1\%)} \\
\bottomrule
\end{tabular}
\end{table}

To understand the \myso dataset, we summarize the content characteristics of the dataset in Table~\ref{tab:dataset_content}.
On the question side, the median token length is 286.5 tokens, but the high standard deviation of 518.3 tokens reflects the wide range of question complexity in Stack Overflow posts, from brief one-line queries to detailed multi-paragraph problem descriptions.
On the answer side, accepted code blocks have a median of 16 lines and 123.5 tokens, confirming that the tasks consist of self-contained, executable snippets rather than skeletal stubs or pseudocode.
The mean TPL import count of 1.6 per task indicates that most tasks involve at least one third-party library, with a non-trivial fraction importing two or more.
Specifically, 46.4\% tasks import at least two distinct libraries and 13.1\% tasks import at least three, providing sufficient multi-dependency structure to observe the version-conflict and concurrent-vulnerability patterns central to RQ2 and RQ3.

\subsection{LLMs Under Evaluation}
\label{sec:setup_models}

The ten LLMs are selected to vary four dimensions simultaneously: deployment prevalence, model scale, knowledge cutoff date, and intra-vendor generational diversity. Table~\ref{tab:models} lists all ten models we evaluated with their basic properties.
Deployment prevalence is represented by the three most widely used commercial coding assistants (GPT-5.4, Claude-Sonnet-4.6, Gemini-3.1-Pro) alongside seven open-source alternatives.
Model scale ranges from 30B to 1T parameters.
Knowledge cutoff spans April 2024 to January 2026.
Intra-vendor diversity is covered by including both Qwen3 and Qwen3.5 from Alibaba to compare successive generations, and Qwen3-235B alongside Qwen3-30B to compare scales within the same generation.
This design enables us to examine whether version-annotation behavior is driven by model capability, training recency, or weight availability, rather than by any single model family.

All models are queried through a unified API under the same calling protocol.
Each query is issued with a timeout of 120 seconds and retried automatically on transient failures to prevent missing data.
The sampling temperature is set to the provider's default to reflect realistic code generation conditions.

\begin{table}[t]\small
\centering
\caption{LLMs evaluated in this study.
  ``Closed'' and ``Open'' refer to model-weight availability.
  For MoE models, active/total parameters are reported.}
\label{tab:models}
\begin{tabular}{lllll}
\toprule
\textbf{Model} & \textbf{Type} & \textbf{Params} & \textbf{Release} & \textbf{Cutoff} \\
\midrule
GPT-5.4                                 & Closed & --         & 2026-03-05 & 2025-08 \\
Claude-Sonnet-4.6                       & Closed & --         & 2026-02-17 & 2025-05 \\
Gemini-3.1-Pro$^{\dag}$                 & Closed & --         & 2026-02-19 & 2025-01 \\
\midrule
DeepSeek-V3.2                           & Open   & 37B/671B   & 2025-12-01 & 2025-05$^{\ddag}$ \\
Kimi-K2.5                               & Open   & 32B/1T     & 2026-01-27 & 2024-04$^{\ddag}$ \\
Qwen3.5-397B$^{\dag}$                   & Open   & 17B/397B   & 2026-02-15 & 2026-01$^{\ddag}$ \\
Qwen3-235B$^{\dag}$                     & Open   & 22B/235B   & 2025-07-21 & 2024-12$^{\ddag}$ \\
Qwen3-30B$^{\dag}$                      & Open   & 3B/30B     & 2025-07-21 & 2024-12$^{\ddag}$ \\
MiniMax-M2.5                            & Open   & 10B/229B   & 2026-02-12 & 2024-06$^{\ddag}$ \\
Llama-4-Scout                           & Open   & 17B/109B   & 2025-04-05 & 2024-08 \\
\bottomrule
\end{tabular}\\
\raggedright\footnotesize
$^{\dag}$Abbreviated from their full names: Gemini-3.1-Pro-Preview, Qwen3.5-397B-A17B, Qwen3-235B-A22B-Instruct-2507, and Qwen3-30B-A3B-Instruct-2507.\\
$^{\ddag}$Knowledge cutoff dates inferred by querying the model directly, as the provider does not officially publish them.
\end{table}

\subsection{Vulnerability Index}
\label{sec:setup_vuln_index}

Stage (4) of the analysis pipeline requires a pre-built mapping from $(\ell_k, v_k)$ pairs to known vulnerability records.
We construct this index offline from the OSV dataset~\cite{osv}, a community-maintained vulnerability database covering the PyPI ecosystem with structured records.

The construction proceeds in three steps.
First, we download all OSV JSON records and retain those whose ecosystem field is \texttt{PyPI}.
Second, for each retained record, we resolve the affected version set from two sources: explicit version lists in the record, and range specifications encoding \texttt{introduced}, \texttt{fixed}, and \texttt{last\_affected} boundaries.
Range specifications are expanded into concrete version enumerations by cross-referencing the full PyPI release history of each package.
Where a single OSV record carries multiple aliases, we deduplicate to the canonical CVE identifier before indexing.
Finally, we build the index mapping each $(\ell_k, v_k)$ pair to its canonical CVE identifiers and their CVSS severity scores for the vulnerability check.

Both the OSV snapshot and the PyPI release metadata are anchored to February 2026, ensuring a consistent vulnerability knowledge base across all pipeline runs.

\subsection{Experiment Environment}
\label{sec:setup_env}

All experiments run on a virtual server with a 16-core AMD EPYC 7713 CPU, 64\,GB RAM, and 1.5\,TB storage, running Ubuntu 24.04 LTS.

\textbf{Version fallback.}
For library references without an explicit version specifier, we resolve the version to the latest PyPI release available as of February 2026, the unified temporal anchor for all models and pipeline runs as discussed in Stage (2) of §\ref{sec:pipeline}.

\textbf{Environment isolation.}
Each $G(t, m, p)$ triple is evaluated in a fully isolated temporary Python virtual environment managed by \texttt{uv@0.10.7}~\cite{uv}, which resolves and installs all dependencies from scratch per run to prevent cross-task package contamination.
We use Python 3.12 as the primary runtime, which was released in October 2023~\cite{py312} and falls within the cutoff of evaluated models.

\textbf{Static type checker.}
We use \texttt{ty@0.0.23}~\cite{ty} to assess static compatibility.
\texttt{ty} resolves type information directly from the packages installed in the \texttt{uv} environment, so it reflects the exact API surface of the specified versions.
We invoke \texttt{ty} with its default configuration and treat any reported diagnostic as an incompatibility signal.

\textbf{BigCodeBench subset.}
BigCodeBench~\cite{bigcodebench} is a large-scale code generation benchmark constructed in Python, containing instruction-following prompts, reference solution code, and test suites.
Of BigCodeBench's 1,140 tasks, 813 contain at least one third-party library import.
We further restrict to the 724 tasks where at least one imported library appears in the vulnerability index, applying the same OSV-membership criterion used for \myso, so that security and compatibility analyses can be applied consistently alongside dynamic execution.
Each task executes within the same isolated \texttt{uv} environment as Stage (5), with a per-task timeout of 180 seconds.


\section{Results}
\label{sec:results}

This section presents the empirical findings of all three research questions.
§\ref{sec:rq1} characterizes whether and how LLMs annotate dependencies with version identifiers, and what those identifiers look like (RQ1).
§\ref{sec:rq2} measures the security consequences of those version choices (RQ2).
§\ref{sec:rq3} evaluates whether the chosen versions are compatible with the generated code (RQ3).

\begin{table*}[t]\small
\centering
\caption{Version specification behavior and vulnerability exposure across models and prompting modes.
  \#LibUses denotes total third-party library mentions across all tasks.
  $\rho_S$(\%) is the proportion of mentions with explicit version declarations.
  $\rho_V$(\%) is the proportion of declared versions resolvable to published PyPI releases.
  $\rho_U$(\%) is the share of resolved versions with at least one known CVE.
  $\rho_H$(\%) reports the fraction of vulnerable versions rated \textit{Critical} or \textit{High}.
  $\tau_U$(\%) is the proportion of tasks exposed to at least one vulnerability.}
\label{tab:rq1_rq2_combined}
\resizebox{\linewidth}{!}{
\begin{tabular}{llrrrrrrr}
\toprule
& \textbf{Model} & \textbf{\#Tasks} & \textbf{\#LibUses} & \textbf{$\rho_S(\%)$} & \textbf{$\rho_V(\%)$} & \textbf{$\rho_U(\%)$} & \textbf{$\rho_H(\%)$} & \textbf{$\tau_U(\%)$} \\
\midrule
\multirow{10}{*}{\rotatebox{90}{\textbf{Explicit}}} & GPT-5.4 & 1000 & 2728 & 6.45 & 98.86 & 21.96 & 310/427(72.60\%) & 28.10 \\
 & Claude-Sonnet-4.6 & 1000 & 2991 & 15.45 & 95.02 & 23.91 & 393/536(73.32\%) & 33.80 \\
 & Gemini-3.1-Pro & 1000 & 2689 & 8.48 & 100.00 & 24.58 & 375/495(75.76\%) & 30.60 \\
\cmidrule[0.5pt](lr){2-9}
 & DeepSeek-V3.2 & 1000 & 3504 & 51.34 & 98.00 & 16.30 & 63/521(12.09\%) & 23.20 \\
 & Kimi-K2.5 & 1000 & 4840 & 45.91 & 98.96 & 26.78 & 808/1024(78.91\%) & 35.60 \\
 & Qwen3.5-397B & 1000 & 2922 & 40.49 & 99.32 & 26.56 & 404/575(70.26\%) & 36.90 \\
 & Qwen3-235B & 1000 & 3104 & 43.49 & 97.33 & 26.90 & 446/610(73.11\%) & 36.90 \\
 & Qwen3-30B & 1000 & 2700 & 59.19 & 91.74 & 32.37 & 474/686(69.10\%) & 44.30 \\
 & MiniMax-M2.5 & 1000 & 2804 & 19.26 & 98.89 & 22.11 & 321/436(73.62\%) & 30.60 \\
 & Llama-4-Scout & 1000 & 2914 & 7.69 & 94.20 & 17.16 & 228/350(65.14\%) & 21.80 \\
\midrule
\multirow{10}{*}{\rotatebox{90}{\textbf{Inline}}} & GPT-5.4 & 1000 & 2387 & 95.18 & 91.29 & 40.51 & 565/827(68.32\%) & 46.20 \\
 & Claude-Sonnet-4.6 & 1000 & 2836 & 88.43 & 85.61 & 38.58 & 588/937(62.75\%) & 50.50 \\
 & Gemini-3.1-Pro & 1000 & 2255 & 91.26 & 86.15 & 42.13 & 568/855(66.43\%) & 50.70 \\
\cmidrule[0.5pt](lr){2-9}
 & DeepSeek-V3.2 & 1000 & 2885 & 70.26 & 85.64 & 40.76 & 699/988(70.75\%) & 54.00 \\
 & Kimi-K2.5 & 1000 & 3532 & 71.89 & 89.88 & 45.78 & 979/1396(70.13\%) & 55.70 \\
 & Qwen3.5-397B & 1000 & 2414 & 94.16 & 83.41 & 40.35 & 515/803(64.13\%) & 50.40 \\
 & Qwen3-235B & 1000 & 2612 & 62.29 & 83.16 & 37.60 & 544/803(67.75\%) & 51.70 \\
 & Qwen3-30B & 1000 & 2676 & 26.83 & 79.39 & 32.85 & 421/565(74.51\%) & 36.70 \\
 & MiniMax-M2.5 & 1000 & 2532 & 91.15 & 87.52 & 45.42 & 730/1028(71.01\%) & 52.10 \\
 & Llama-4-Scout & 1000 & 2866 & 57.92 & 81.81 & 38.59 & 677/954(70.96\%) & 51.70 \\
\bottomrule
\end{tabular}
}
\end{table*}

\subsection{RQ1: Version Specification Behavior}
\label{sec:rq1}

RQ1 asks whether LLMs annotate dependencies with version identifiers and whether those identifiers correspond to valid published releases.
To answer this, we examine four aspects in turn: specification rates across models and modes, the concentration of version choices around particular releases, the validity of specified identifiers, and the temporal lag between specified versions and each model's knowledge cutoff.

\subsubsection{Specification Rate}
\label{sec:rq1_spec}

\textbf{Prompting mode, rather than model capability, is the dominant driver of version-annotation behavior.}
As shown in Table~\ref{tab:rq1_rq2_combined}, under inline mode, seven of the ten models specify versions for more than 70\% of all library uses, with the highest-annotating model reaching 95.18\%.
Qwen3-30B is the sole outlier, annotating only 26.83\% of library uses despite belonging to the same model family as peers that annotate more than 60\% of their references.
The within-family contrast is striking: the two Qwen3 generations from the same vendor differ by nearly 32 percentage points in inline annotation rate, a gap far larger than what parameter count alone can explain.
This suggests that generation-level training differences outweigh raw scale in determining annotation frequency.

Under explicit mode, the same models largely abandon version constraints.
The three closed-source models annotate fewer than 16\% of library uses, with the lowest reaching just 6.45\% as shown in Table~\ref{tab:rq1_rq2_combined}.
When generating a \texttt{requirements.txt} file, these models routinely produce bare library names with no version specifier, directly undermining dependency reproducibility.
Open-source models show wider variation under explicit mode, ranging from 7.69\% to 59.19\%.
Notably, Qwen3-30B inverts the pattern seen in every other model: it achieves the highest explicit-mode rate while producing the lowest inline rate, indicating that its annotation behavior is sensitive to output format in the opposite direction from its peers.
Within the Qwen family, this inversion is progressive: the two larger models each see their annotation rate fall by 20--50 percentage points when switching from inline to explicit mode, while Qwen3-30B shows the opposite, rising by more than 32 points.
The smallest model thus behaves more like a structured manifest generator than an inline annotator.
This behavioral divergence does not obviously track model capacity.

\begin{figure}[t]
  \centering
  \includegraphics[width=1.0\linewidth]{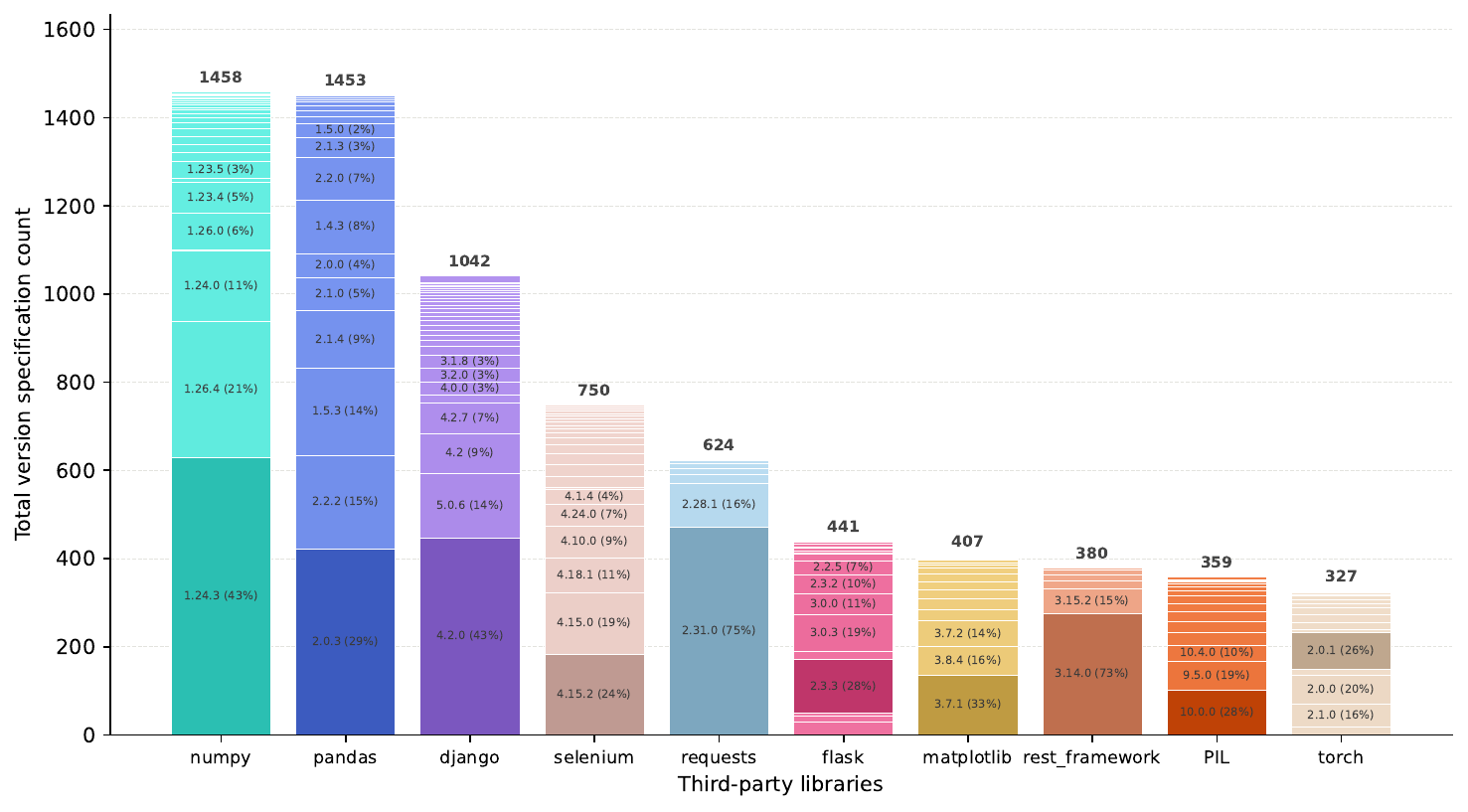}
  \Description{A stacked bar chart showing the distribution of version annotations across the ten most frequently annotated third-party libraries under inline prompting mode, aggregated across all ten evaluated models. Each bar represents a library, with colored segments indicating individual version shares. The chart reveals heavy concentration: for requests, a single version accounts for approximately 75 percent of all annotations; for numpy and django, one version accounts for roughly 43 percent each; for djangorestframework, one version accounts for 73 percent. This pattern demonstrates that LLMs converge on a very small number of dominant versions per library.}
  \caption{Specified version distribution for the top-10 most frequently annotated TPLs under inline mode.
    Each bar shows the total annotation count, and color segments represent individual version shares.
    Version labels are shown for segments exceeding 2\% share.}
  \label{fig:version_concentration}
\end{figure}

Based on the results, these within- and across-family patterns point to the same conclusion that format affordance, not model capability, determines whether versions are annotated.
The inline-explicit gap reaches as wide as 88.73 percentage points for GPT-5.4.
When the output format makes version annotation syntactically natural, most models comply at high rates.
When the format permits omitting versions, the same models largely do so.
Projects that rely on LLM-generated dependency manifests in explicit mode cannot assume that version information will be present.

\subsubsection{Version Concentration}
\label{sec:rq1_concentration}

\textbf{LLMs converge on a small number of versions per library, amplifying any security risk those versions carry.}
Figure~\ref{fig:version_concentration} shows the distribution of inline-mode version annotations for the ten most frequently annotated TPLs, aggregated across all models.
Across these libraries, version choices are heavily concentrated: \texttt{1.24.3} accounts for 43\% of all \texttt{numpy} annotations, 43\% of LLMs annotate \texttt{django} with \texttt{4.2.0}, and 75\% of \texttt{requests} were labeled as \texttt{2.31.0}.
The \texttt{djangorestframework} (\texttt{rest\_framework}) distribution is even more extreme, with \texttt{3.14.0} accounting for 73\% of all annotations for that library.
This convergence is not explained by a lack of available releases: \texttt{requests} alone has over 50 published versions on PyPI, yet models overwhelmingly select a single one.

This raises security implications that if a dominant version carries known CVEs, a large fraction of all LLM-generated code referencing that library inherits the same vulnerability, with no per-task variation.
The convergence is thus not merely a stylistic observation but a structural risk amplifier.

\subsubsection{Validity of Specified Versions}
\label{sec:rq1_validity}

\textbf{Most inline-specified versions are valid PyPI releases, but a consistent invalidity rate of 7--16\% persists across all models.}
Across 26,691 inline library uses, 2,998 version strings (11.23\%) resolve to no published PyPI release.
At the model level, Qwen3-30B has the lowest invalidity rate of 7.32\% and Qwen3.5-397B the highest of 16.28\%.
Within the Qwen family, inline validity is remarkably stable despite large differences in specification rate: all three Qwen models cluster within four percentage points of each other in validity, confirming that version-knowledge accuracy is not strongly coupled to annotation frequency.
Under explicit mode, validity rates are uniformly higher, at 91.74\%--100.00\%, but this partly reflects the version-resolution mechanism: library uses without a specifier are resolved to the latest PyPI release, which is valid by construction.
Inline validity therefore constitutes the more informative measure of model-level version knowledge.

Inspection of the 2,998 invalid strings reveals two structural categories.
\emph{Version hallucination} counts for 91.96\% of invalidity at 2,757 cases: a version string that follows the library's versioning convention but was never released.
In such cases, models generate version identifiers as if they were valid releases.
A recurring example is \texttt{opencv-python==4.8.0}, which appears 16 times each in MiniMax-M2.5 and Qwen3-235B output and 14 times in Qwen3.5-397B output.
The underlying error is a suffix omission: OpenCV uses four-part version identifiers, \textit{e.g.}, \texttt{4.8.0.74}, \texttt{4.8.0.76}, but models consistently generate the truncated form \texttt{4.8.0}, which does not exist as a published release.
This suffix-omission pattern is consistent across model families, indicating that it reflects a shared training-signal artifact rather than any individual model's idiosyncrasy.
The second category is \emph{cross-library confusion} that has 241 cases (8.04\%), where the resolved package name is not found on PyPI because the import name and the package name diverge. Models use the Python import identifier as a PyPI package name.
Examples include \texttt{scikit\_learn} for \texttt{scikit-learn}, \texttt{azure\_storage} for \texttt{azure-storage-blob}, and \texttt{bio} for \texttt{biopython}.
Some of these failures are attributable to gaps in the \texttt{pipreqs} import-to-package mapping used in Stage (2) of §\ref{sec:pipeline}: packages not covered by the mapping table are resolved by their raw import name, which may differ from their PyPI name.

\subsubsection{Version Recency}
\label{sec:rq1_temporal}


\begin{figure}[t]
  \centering
  \includegraphics[width=1.0\linewidth]{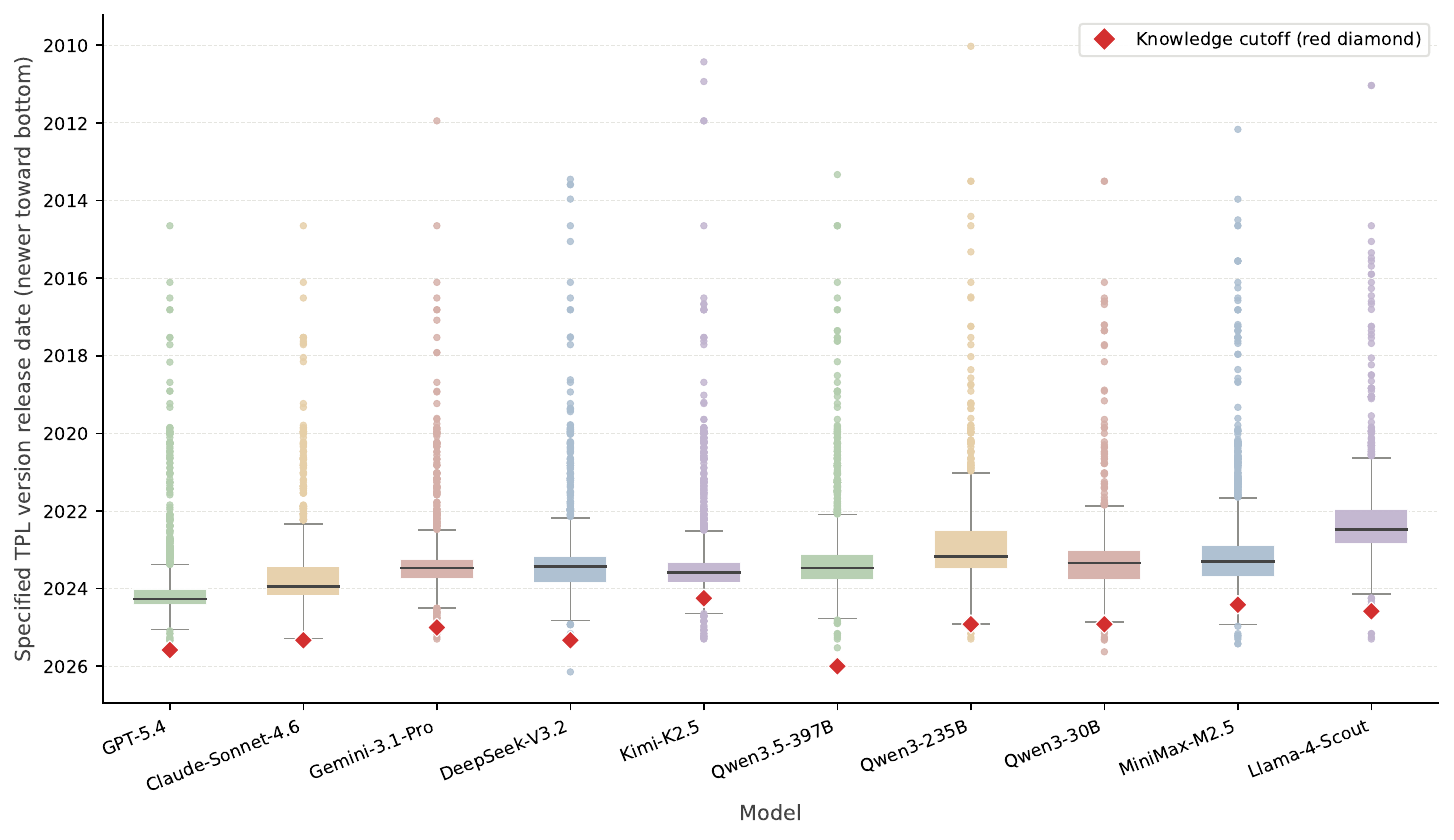}
  \Description{A box plot showing the distribution of release dates of LLM-specified library versions per model under inline prompting mode. The y-axis is inverted so that newer release dates appear toward the bottom. Red diamond markers indicate each model's knowledge cutoff date. For all ten models, the median of the specified version release date distribution lies above the cutoff diamond, indicating that models systematically prefer versions released well before their training cutoff. The median lag ranges from 9 months (Kimi-K2.5) to 31 months (Qwen3.5-397B).}
  \caption{Distribution of release dates of LLM-specified TPL versions per model under inline mode.
    The newer dates appear toward the bottom.
    Red diamonds mark each model's knowledge cutoff date.
    For all ten models, the median specified version release date falls above (older than) the cutoff diamond.}
  \label{fig:version_release_time}
\end{figure}

\textbf{All ten models specify versions that substantially predate their knowledge cutoffs, with median lags ranging from 9 to 31 months.}
Figure~\ref{fig:version_release_time} shows the release-date distribution of inline-mode specified versions per model.
Kimi-K2.5 has the shortest lag: its median specified version was released in July 2023, nine months before its April 2024 knowledge cutoff.
Several other models fall in the 14--17 month range. For example, GPT-5.4's median specified version predates its knowledge cutoff by approximately 16 months.
At the far end, Qwen3.5-397B has a 31-month lag: its knowledge cutoff is January 2026, yet its median specified version was released in June 2023.
Parametric recency does not translate into version recency: despite carrying the most recent training cutoff among all evaluated models, Qwen3.5-397B does not produce the most recent version choices.

The Qwen3 within-family comparison is equally instructive.
Qwen3-235B and Qwen3-30B share an identical December 2024 cutoff, yet their median release dates are March 2023 and May 2023, yielding lags of 21 and 19 months respectively.
Their distributions are similar despite a 7$\times$ difference in active parameter count.
The newer-generation Qwen3.5-397B, by contrast, lags its own cutoff by 31 months, substantially more than either Qwen3 model, confirming that the lag is shaped by training data composition rather than model scale or recency.

Several models produce outlier versions released after their own knowledge cutoffs.
GPT-5.4 and Qwen3.5-397B are the only exceptions with zero post-cutoff versions.
Among the remaining eight models, post-cutoff rates range from near zero to 3.75\%, with Kimi-K2.5 being the most frequent extrapolator at 95 post-cutoff version strings.
The most widespread post-cutoff string is \texttt{numpy==2.2.5}, released April 2025, which appears across six models whose cutoffs range from April 2024 to January 2025.
These strings cannot have been learned from training data.
These models are extrapolating plausible next version numbers for packages they have never seen released.
That the extrapolations converge on the same high-profile release suggests models infer near-future version numbers from observed versioning patterns rather than generating them arbitrarily.

In summary, the median lags of 9 to 31 months across all models are consistent with a training-signal hypothesis: a version released years before the cutoff accumulates far more representation in documentation and dependency files than a recently released version, biasing LLM choices toward a popular version window that substantially precedes the release frontier.

\begin{summary}
\textbf{Summary:} Under inline prompting, LLMs specify version identifiers for 26.83\%--95.18\% of library references; under explicit prompting, the same models specify only 6.45\%--59.19\%, with the gap driven by format affordance rather than consistent engineering intent.
Among inline-specified versions, 79.39\%--91.29\% correspond to valid PyPI releases, with the remainder attributable primarily to version hallucination.
Annotations concentrate around a small number of dominant versions per library.
Those versions systematically lag each model's knowledge cutoff by 9 to 31 months, reflecting uneven representation of version-specific data in model training.
\end{summary}

\subsection{RQ2: Version Vulnerability Exposure}
\label{sec:rq2}

Having established that LLMs specify dependency versions at high rates under inline prompting and that most specified versions are valid PyPI releases, we now ask what the security implications of those choices are.
RQ2 measures how frequently the specified versions carry known CVE vulnerabilities and characterizes the severity, structural origins, and temporal properties of that exposure.
We focus on inline mode throughout this section. As §\ref{sec:rq1} showed, explicit mode produces so few version annotations that model-level vulnerability comparisons are less informative.

\subsubsection{Overall Vulnerability Exposure}
\label{sec:rq2_overall}

\begin{figure}[t]
  \centering
  \includegraphics[width=\linewidth]{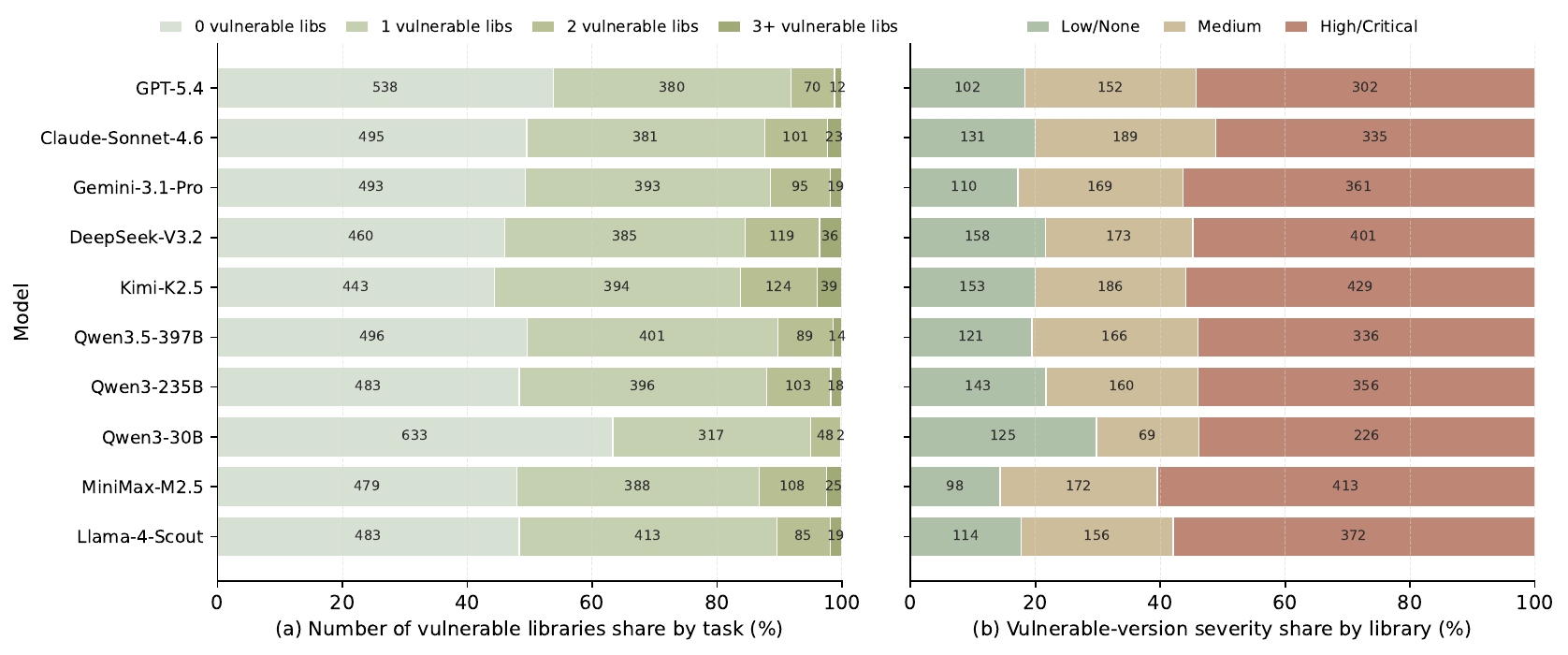}
  \Description{Two subfigures showing per-model vulnerability exposure under inline prompting mode. Subfigure (a) shows the distribution of the number of vulnerable library versions per task for each model, revealing that multi-library vulnerability is common among exposed tasks. Subfigure (b) shows the severity composition of all vulnerable versions specified by each model, broken down by CVSS severity levels: None, Low, Medium, High, and Critical. Critical and High severity CVEs account for 62.75 to 74.51 percent of all vulnerable versions across models, substantially exceeding the baseline distribution in the OSV dataset.}
  \caption{Per-model vulnerability exposure profile under inline mode.
    (a): Distribution of the number of vulnerable library versions per task.
    (b): Severity composition (\textit{None}/\textit{Low}/\textit{Medium}/\textit{High}/\textit{Critical}) of all vulnerable versions specified by each model.}
  \label{fig:vuln_exposure_severity}
\end{figure}

\textbf{LLMs expose 36.70\%--55.70\% of tasks to known vulnerabilities through their version choices.}
As shown in Table~\ref{tab:rq1_rq2_combined}, task-level vulnerability exposure under inline prompting ranges from 36.70\% for Qwen3-30B to 55.70\% for Kimi-K2.5.
Figure~\ref{fig:vuln_exposure_severity}(a) shows that multi-library vulnerability is common among exposed tasks.
For Kimi-K2.5, the model with the highest exposure, nearly 30\% of its exposed tasks contain two or more vulnerable library versions simultaneously.
For Qwen3-30B, the model with the lowest exposure is around 14\%.
Vulnerability exposure is therefore not limited to isolated single-library incidents but frequently involves concurrent vulnerabilities across multiple dependencies in the same generated snippet.

\textbf{High-severity vulnerabilities dominate across all models.}
As shown in Figure~\ref{fig:vuln_exposure_severity}(b) and Table~\ref{tab:rq1_rq2_combined}, \textit{Critical} and \textit{High} severity CVEs account for 62.75\%--74.51\% of all vulnerable versions.
For context, the OSV dataset baseline distributes PyPI vulnerability severity as 47.86\% Low/None, 23.24\% Medium, and 28.90\% High/Critical across all known records, without filtering by LLM preference.
LLM-recommended versions therefore carry a disproportionate share of high-severity CVEs relative to the overall vulnerability landscape.
This overrepresentation suggests that the versions LLMs prefer are not merely old, but are specifically those that accumulated the most severe unpatched vulnerabilities before fixes became widely adopted.

\subsubsection{Vulnerable Version Convergence}
\label{sec:rq2_convergence}

\begin{figure}[t]
  \centering
  \includegraphics[width=\linewidth]{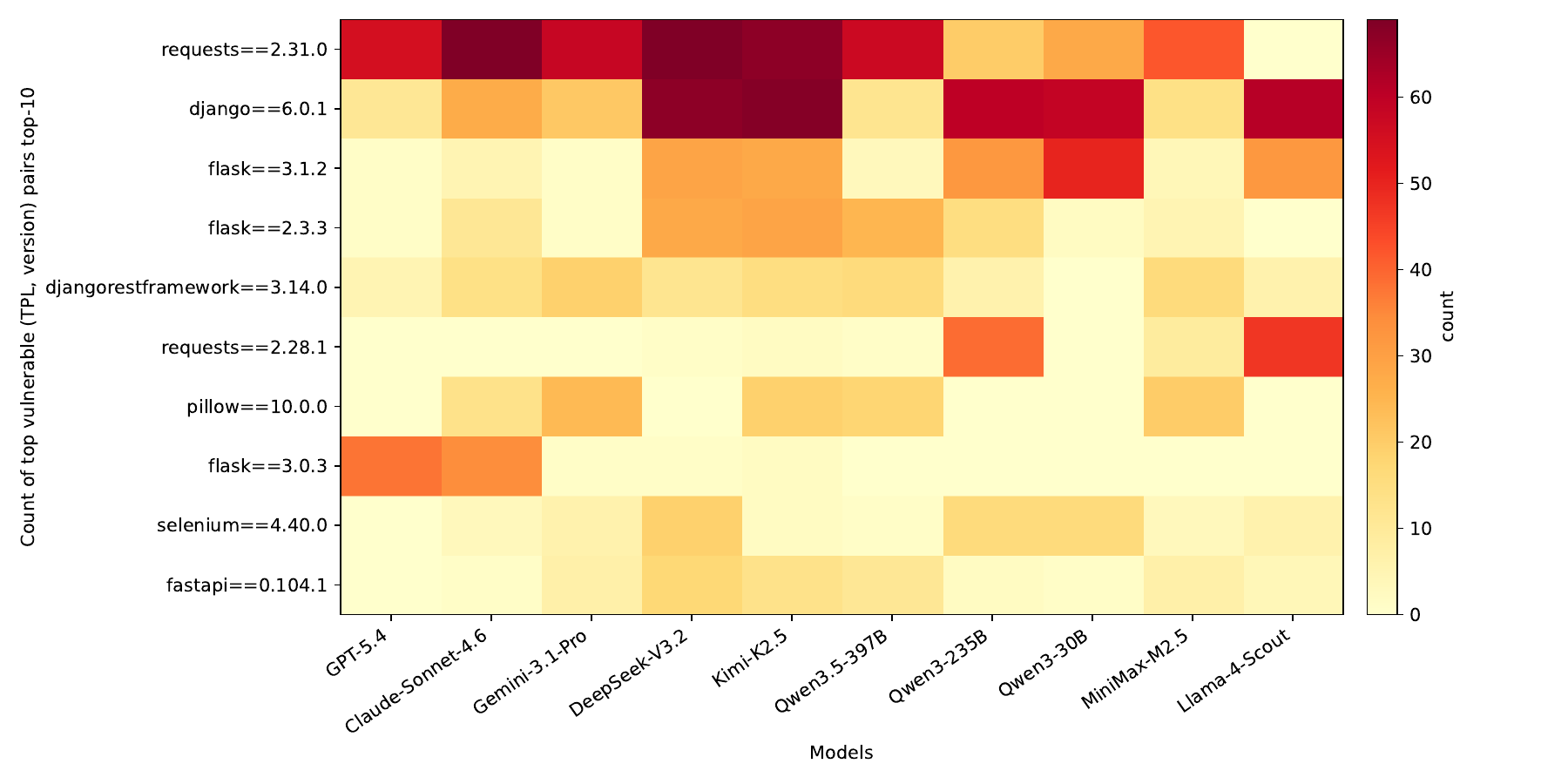}
  \Description{A heatmap showing cross-model convergence on the top-10 most frequently specified vulnerable library-version pairs under inline prompting mode. Rows correspond to vulnerable pairs (e.g., requests==2.31.0, django==6.0.1, flask==3.1.2) sorted by total specification count across all models; columns correspond to the ten evaluated models. Cell color encodes specification count, with darker cells indicating higher counts. Rows with uniform coloring across all columns indicate cross-model consensus; rows with color concentrated in one column indicate model-specific preference. The top five rows are nearly uniformly colored, confirming that the most common vulnerable versions are shared across all model families.}
  \caption{Cross-model convergence on the most frequently specified vulnerable
    $(\ell,v)$ pairs under inline mode.
    Cell color encodes specification count; rows are sorted by total count across all models.
    Uniform row coloring indicates cross-model consensus on the same vulnerable version.
    Concentrated single-column coloring indicates model-specific preference.}
  \label{fig:vuln_convergence}
\end{figure}

\textbf{Vulnerable version choices are shared across model families, not idiosyncratic to individual models.}
Figure~\ref{fig:vuln_convergence} plots the top-10 most frequently specified vulnerable $(\ell,v)$ pairs as a heatmap across all ten models.
Across 6,378 total vulnerable version assignments spanning 1,289 unique vulnerable $(\ell,v)$ pairs, the top-10 entries alone account for 26.34\% of all assignments.
The top five entries are each specified by at least nine of the ten models, while seven of the ten top entries meet this threshold.
The single most convergent pair is \texttt{django==6.0.1}, specified 400 times across all ten models.

Uniform row coloring is visible for the top entries, confirming that convergence rather than model-specific preference drives the most common vulnerable choices.
The lower entries show more model-specific concentration, but the five most convergent pairs, \texttt{requests==2.31.0}, \texttt{django==6.0.1}, \texttt{flask==3.1.2}, \texttt{flask==2.3.3}, and \texttt{djangorestframework==3.14.0}, each appear in at least nine of the ten models' outputs, spanning both closed-source and open-source families.
This breadth rules out any single training pipeline as the source of the bias.
The pattern directly corroborates the version-concentration finding in §\ref{sec:rq1_concentration}: the same narrow set of popular releases that LLMs overwhelmingly select also happens to carry the most prevalent CVEs, confirming that high version concentration translates directly into concentrated vulnerability exposure.

\subsubsection{Temporal Awareness of Known Vulnerabilities}
\label{sec:rq2_temporal}

The convergence result raises a further question: were the CVEs carried by LLM-specified versions publicly disclosed before training, making the exposure in principle avoidable, or only disclosed after?

\begin{figure}[t]
  \centering
  \includegraphics[width=0.84\linewidth]{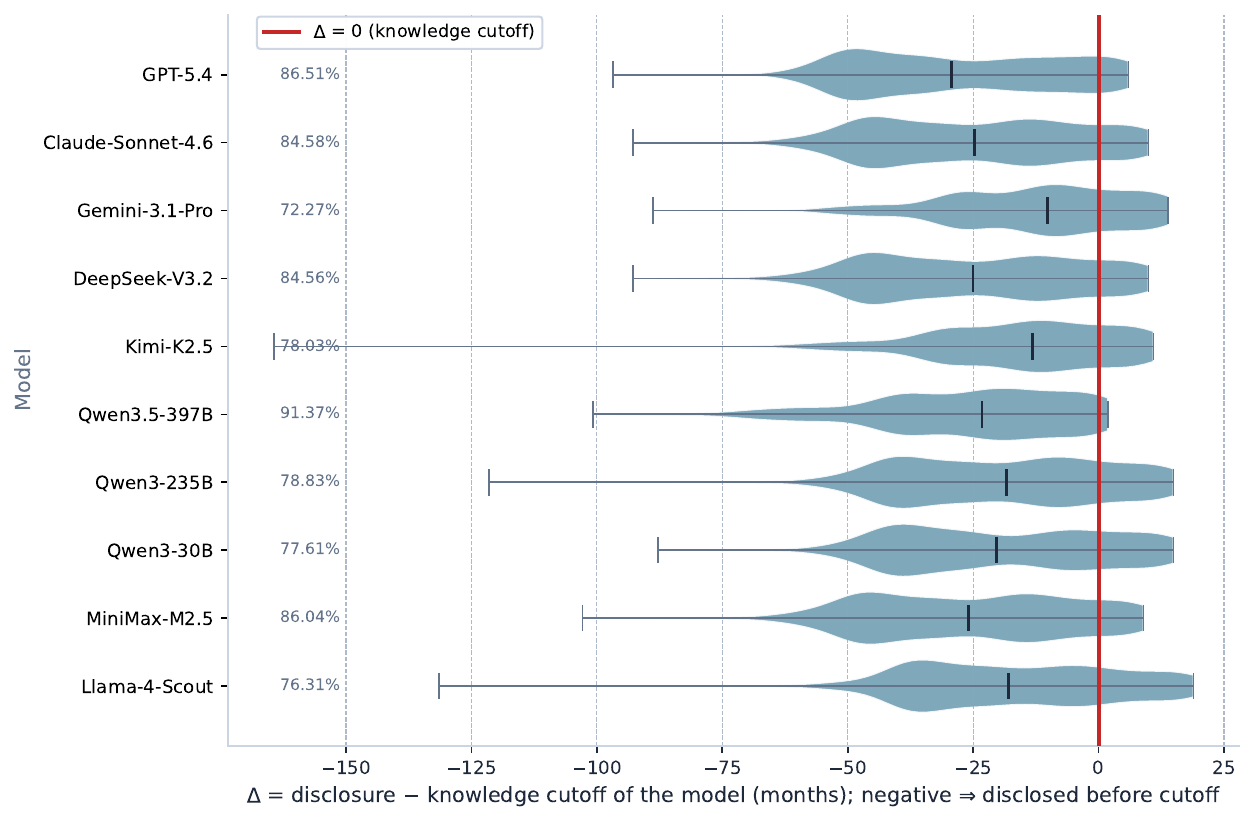}
  \Description{A violin plot showing the distribution of the lag between CVE disclosure dates and each model's knowledge cutoff, defined as Delta equals disclosure date minus cutoff date in months, for all vulnerabilities carried by LLM-specified versions under inline prompting mode. Negative Delta values indicate that the CVE was publicly disclosed before the model's training cutoff. A red vertical line marks Delta equals zero. For all ten models, the bulk of each distribution lies to the left of zero, with long left tails extending to approximately negative 150 months. This demonstrates that the vast majority of CVEs carried by LLM-specified versions were already part of the public record at training time, making the exposure in principle avoidable.}
  \caption{Distribution of the lag between CVE disclosure dates and each model's
    knowledge cutoff ($\Delta = \text{disclosure} - \text{cutoff}$, in months)
    for all vulnerabilities carried by LLM-specified versions under inline mode.
    Negative values indicate that the CVE was publicly disclosed before the
    model's training cutoff. The red line marks $\Delta = 0$.
    The bulk of each distribution lies left of zero, showing that the vast
    majority of vulnerabilities in LLM-recommended versions were already part
    of the public record at training time.}
  \label{fig:cve_disclosure}
\end{figure}

\textbf{Most CVEs carried by LLM-specified versions were publicly disclosed before each model's training cutoff.}
Figure~\ref{fig:cve_disclosure} plots $\Delta = \text{disclosure} - \text{cutoff}$ in months.
Negative values indicate pre-cutoff disclosure.
As shown in Figure~\ref{fig:cve_disclosure}, across all 1,410 unique CVE identifiers observed in the study, 72.27\%--91.37\% of the unique CVEs associated with each model's specified versions were disclosed before that model's knowledge cutoff.
The proportion is highest for Qwen3.5-397B at 91.37\%, which carries the most recent knowledge cutoff of January 2026, and lowest for Gemini-3.1-Pro at 72.27\%, which has an earlier cutoff of January 2025.
Aggregated across all models, 81.8\% of unique CVE-model pairs correspond to vulnerabilities that were part of the public record at training time.

For all ten models, the bulk of mass in Figure~\ref{fig:cve_disclosure} lies left of $\Delta = 0$, with long left tails extending to $\Delta \approx -150$ months.
Several outlier CVEs dating to 2007--2008 are visible at the extreme left.
These are decade-old vulnerabilities still carried by widely-used library versions, present in LLM-recommended code because those versions accumulated extraordinary web presence long before the vulnerabilities were well-known.
LLM-introduced vulnerability exposure is therefore not primarily a consequence of training-data latency.
The relevant CVEs were accessible during training, making the exposure an avoidable rather than an inherent limitation.

\subsubsection{Availability of Safe Alternatives}
\label{sec:rq2_safe}

\textbf{Version-level vulnerability exposure cannot be corrected by minor version substitution.}
The findings above establish that LLM-specified versions frequently carry known CVEs.
A natural question is whether exposure could be resolved by choosing a patch version of the same library.
We examine whether, for each LLM-specified vulnerable $(\ell,v)$ pair, a CVE-free alternative exists within the same major version series.
Such a substitution would be minimal, preserving API compatibility in most cases.
Out of 1,289 unique vulnerable $(\ell,v)$ pairs identified across all models, only 63 have at least one CVE-free release in the same major version branch.
For 95.11\% of vulnerable pairs, the entire same-major version branch is affected by at least one CVE, leaving no safe same-major alternative available.
Developers who discover a vulnerable LLM-specified version cannot simply increment the patch number.
They face a cross-major migration or a deliberate security-aware version selection process.
This structural characteristic makes the problem qualitatively different from ordinary version drift and underscores the need for external tooling rather than local version adjustment.

\begin{summary}
\textbf{Summary:} Under inline prompting, 36.70\%--55.70\% of tasks contain at least one LLM-specified library version carrying a known CVE, with 62.75\%--74.51\% of vulnerable versions rated \textit{Critical} or \textit{High} severity.
The most frequently specified vulnerable versions are shared across all models, indicating a systematic bias rather than model-specific error.
The associated CVEs were publicly disclosed before each model's knowledge cutoff in 72\%--91\% of cases, making the exposure avoidable in principle.
Finally, 95.11\% of vulnerable $(\ell,v)$ pairs have no CVE-free alternative within the same major version branch, showing a structural problem.
\end{summary}

\subsection{RQ3: Version-Code Compatibility}
\label{sec:rq3}

\begin{table*}[t]\small
\centering
\caption{Static compatibility rates and quadrant task breakdown.
  $\tau_C(\%)$ is the passing rate for static type check.
  Quadrant entries partition tasks by the intersection of compatibility and vulnerability status.}
\label{tab:pipeline_d2_compat_quadrants}
\resizebox{\linewidth}{!}{%
\begin{tabular}{llrrrrr}
\toprule
& \textbf{Model} & \textbf{$\tau_C(\%)$} & \textbf{Safe$\cap$Compat} & \textbf{Unsafe$\cap$Compat} & \textbf{Safe$\cap$Incompat} & \textbf{Unsafe$\cap$Incompat} \\
\midrule
\multirow{10}{*}{\rotatebox{90}{\textbf{Explicit}}} & GPT-5.4 & 63.20 & 49.30 & 13.90 & 22.60 & 14.20 \\
 & Claude-Sonnet-4.6 & 53.40 & 40.50 & 12.90 & 25.70 & 20.90 \\
 & Gemini-3.1-Pro & 66.70 & 49.60 & 17.10 & 19.80 & 13.50 \\
\cmidrule[0.5pt](lr){2-7}
 & DeepSeek-V3.2 & 48.50 & 40.60 & 7.90 & 36.20 & 15.30 \\
 & Kimi-K2.5 & 51.90 & 37.90 & 14.00 & 26.50 & 21.60 \\
 & Qwen3.5-397B & 62.30 & 42.60 & 19.70 & 20.50 & 17.20 \\
 & Qwen3-235B & 52.20 & 38.20 & 14.00 & 24.90 & 22.90 \\
 & Qwen3-30B & 45.90 & 31.20 & 14.70 & 24.50 & 29.60 \\
 & MiniMax-M2.5 & 55.20 & 42.00 & 13.20 & 27.40 & 17.40 \\
 & Llama-4-Scout & 45.30 & 37.60 & 7.70 & 40.60 & 14.10 \\
\midrule
\multirow{10}{*}{\rotatebox{90}{\textbf{Inline}}} & GPT-5.4 & 63.20 & 36.90 & 26.30 & 16.90 & 19.90 \\
 & Claude-Sonnet-4.6 & 46.00 & 27.20 & 18.80 & 22.30 & 31.70 \\
 & Gemini-3.1-Pro & 50.60 & 26.50 & 24.10 & 22.80 & 26.60 \\
\cmidrule[0.5pt](lr){2-7}
 & DeepSeek-V3.2 & 37.60 & 19.80 & 17.80 & 26.20 & 36.20 \\
 & Kimi-K2.5 & 41.80 & 20.90 & 20.90 & 23.40 & 34.80 \\
 & Qwen3.5-397B & 46.10 & 23.00 & 23.10 & 26.60 & 27.30 \\
 & Qwen3-235B & 34.40 & 19.30 & 15.10 & 29.00 & 36.60 \\
 & Qwen3-30B & 44.60 & 32.10 & 12.50 & 31.20 & 24.20 \\
 & MiniMax-M2.5 & 19.70 & 12.10 & 7.60 & 35.80 & 44.50 \\
 & Llama-4-Scout & 25.40 & 15.20 & 10.20 & 33.10 & 41.50 \\
\bottomrule
\end{tabular}%
}
\end{table*}

RQ1 and RQ2 established that LLMs frequently pin specific library versions and that those versions often carry known CVEs.
RQ3 asks a complementary question.
Even setting security aside, are the pinned versions actually compatible with the generated code?
We assess this through two stages: static compatibility, assessed via installation verification and \texttt{ty} static type checking, and dynamic compatibility, assessed via BigCodeBench test suite execution.

\subsubsection{Static Compatibility}
\label{sec:rq3_static}

\textbf{Inline prompting produces substantially lower static compatibility rates than explicit prompting.}
This difference reflects the distinct version-resolution mechanisms of the two modes.
Table~\ref{tab:pipeline_d2_compat_quadrants} reports per-model compatibility rates.
Under explicit prompting, rates range from 45.30\% to 66.70\%.
Because most models annotate fewer than 20\% of library references in explicit mode, the majority of dependencies are resolved to their latest PyPI release at the study's temporal anchor.
Latest releases install successfully by construction, inflating the apparent compatibility rate.
Under inline prompting, where models pin specific versions on every import, compatibility rates fall considerably, ranging from 19.70\% to 63.20\% across models.
Most inline models fall below 50\%.
GPT-5.4 is the only model that achieves equal compatibility across both modes, at 63.20\%.

The \textbf{Unsafe\,$\cap$\,Incompat} quadrant captures tasks that simultaneously expose the user to a known CVE and fail static checking.
Under inline prompting, this quadrant accounts for 19.90\% to 44.50\% of tasks across models.
Only GPT-5.4 achieves a \textbf{Safe\,$\cap$\,Compatible} rate above 30\%, at 36.90\%.
All other inline models fall below 27\%.
Fully safe and executable code is therefore the exception rather than the norm under inline version pinning.

\begin{table*}[t]\small
\centering
\caption{Static compatibility check failure diagnostics.
  For each incompatible task, the first reported \texttt{ty} check rule is used as its primary diagnostic label.
  Columns show the six globally most frequent error categories plus a catch-all.
  \textbf{InstErr}: dependency installation failure,
  \textbf{UnrAttr}: unresolved attribute reference,
  \textbf{UnrImp}: unresolved import,
  \textbf{InvSyn}: invalid syntax,
  \textbf{InvArg}: invalid argument type,
  \textbf{UnrRef}: unresolved reference.}
\label{tab:pipeline_d2_ty_error_overview}
\resizebox{\linewidth}{!}{%
\begin{tabular}{llrrrrrrrr}
\toprule
 & \textbf{Model} & \textbf{InstErr} & \textbf{UnrAttr} & \textbf{UnrImp} & \textbf{InvSyn} & \textbf{InvArg} & \textbf{UnrRef} & \textbf{Others} & \textbf{Total} \\
\midrule
\multirow{10}{*}{\rotatebox{90}{\textbf{Explicit}}} & GPT-5.4 & 129 & 56 & 37 & 25 & 26 & 7 & 88 & 368 \\
 & Claude-Sonnet-4.6 & 221 & 70 & 57 & 8 & 23 & 9 & 78 & 466 \\
 & Gemini-3.1-Pro & 118 & 70 & 40 & 5 & 19 & 1 & 80 & 333 \\
\cmidrule[0.5pt](lr){2-10}
 & DeepSeek-V3.2 & 309 & 51 & 40 & 9 & 24 & 9 & 73 & 515 \\
 & Kimi-K2.5 & 219 & 60 & 43 & 34 & 34 & 17 & 74 & 481 \\
 & Qwen3.5-397B & 174 & 52 & 49 & 4 & 22 & 3 & 73 & 377 \\
 & Qwen3-235B & 229 & 58 & 57 & 11 & 30 & 15 & 78 & 478 \\
 & Qwen3-30B & 327 & 51 & 58 & 10 & 15 & 28 & 52 & 541 \\
 & MiniMax-M2.5 & 136 & 60 & 70 & 50 & 21 & 28 & 83 & 448 \\
 & Llama-4-Scout & 204 & 73 & 109 & 10 & 29 & 23 & 99 & 547 \\
\midrule
\multirow{10}{*}{\rotatebox{90}{\textbf{Inline}}} & GPT-5.4 & 124 & 58 & 39 & 2 & 55 & 11 & 79 & 368 \\
 & Claude-Sonnet-4.6 & 241 & 44 & 40 & 126 & 20 & 14 & 55 & 540 \\
 & Gemini-3.1-Pro & 343 & 51 & 31 & 0 & 14 & 2 & 53 & 494 \\
\cmidrule[0.5pt](lr){2-10}
 & DeepSeek-V3.2 & 462 & 39 & 46 & 11 & 18 & 8 & 40 & 624 \\
 & Kimi-K2.5 & 406 & 45 & 42 & 7 & 18 & 10 & 54 & 582 \\
 & Qwen3.5-397B & 388 & 37 & 54 & 3 & 11 & 5 & 41 & 539 \\
 & Qwen3-235B & 472 & 50 & 57 & 7 & 14 & 12 & 44 & 656 \\
 & Qwen3-30B & 253 & 57 & 98 & 14 & 21 & 41 & 70 & 554 \\
 & MiniMax-M2.5 & 400 & 17 & 18 & 337 & 5 & 1 & 25 & 803 \\
 & Llama-4-Scout & 452 & 70 & 73 & 35 & 14 & 54 & 48 & 746 \\
\bottomrule
\end{tabular}%
}
\end{table*}

\subsubsection{Compatibility Error Analysis}
\label{sec:rq3_errors}

\textbf{Installation failure is the dominant cause of static incompatibility under inline prompting.}
Table~\ref{tab:pipeline_d2_ty_error_overview} reports the primary diagnostic for each incompatible task.
Every run in each task of \myso is processed in a new isolated temporary Python venv, we install the LLM specified TPLs at their versions and run the corresponding checks as illustrated in \S\ref{sec:pipeline}.
Under inline mode, dependency installation error is the single largest category for every model, ranging from 124 tasks for GPT-5.4 to 472 for Qwen3-235B, showing LLMs specify versions that are not even installable.
API-level diagnostics reported after a successful installation, namely \texttt{unresolved-attribute}, \texttt{unresolved-import}, and \texttt{invalid-argument-type}, reported as UnrAttr, UnrImp, and InvArg in Table~\ref{tab:pipeline_d2_ty_error_overview}, together account for a smaller but consistent share.
These reflect cases where the dependency set installs, but the generated code references API surfaces that do not exist in the pinned version. This shows LLMs cannot fully correspond to the API usage of TPLs with their version.

Among installation failures, dependency resolution conflict is the most prevalent sub-cause across all models, accounting for 84--249 cases per model under inline prompting.
These conflicts arise when pinned versions impose mutually incompatible transitive dependency constraints that the resolver cannot satisfy simultaneously.
A distinct cluster, \texttt{missing\_distutils}, accounts for 8--134 inline failures per model.
It reflects a systematic version-era mismatch: packages whose metadata was written before Python removed \texttt{distutils} from the standard library in version 3.12 fail to install in the evaluation environment.
This pattern is a direct consequence of the temporal lag documented in §\ref{sec:rq1_temporal}: LLMs preferentially specify older versions, and those versions disproportionately carry dependencies on a module that no longer exists.
Under explicit prompting, \texttt{missing\_distutils} failures are far rarer, as latest-version fallback resolution tends to select releases that have already dropped the deprecated dependency.

Two models exhibit anomalously high \texttt{invalid-syntax} counts under inline mode.
MiniMax-M2.5 accounts for 337 such tasks and Claude-Sonnet-4.6 for 126.
This pattern is distinct from the version-level failures that dominate elsewhere.
It indicates that these models occasionally generate syntactically malformed code when producing inline version annotations, introducing a code-quality failure independent of any version choice.
Taken together, the error taxonomy shows that version selection, not code logic, accounts for the overwhelming share of incompatibility failures under inline mode.

\subsubsection{Dynamic Compatibility}
\label{sec:rq3_dynamic}

\begin{table*}[t]\small
\centering
\caption{Measurement results of BigCodeBench test case execution.
\#Pass, \#Fail, and \#Error report task-level test outcomes, 
\#Total denotes the number of measured tasks, 
and PassRate(\%) is the pass rate among them.}
\label{tab:pipeline_d1_bcb_results_osv_subset}
\begin{tabular}{llrrrrr}
\toprule
& \textbf{Model} & \textbf{\#Pass} & \textbf{\#Fail} & \textbf{\#Error} & \textbf{\#Total} & \textbf{PassRate(\%)} \\
\midrule
\multirow{10}{*}{\rotatebox{90}{\textbf{Explicit}}} & GPT-5.4 & 347 & 372 & 5 & 724 & 47.93 \\
 & Claude-Sonnet-4.6 & 353 & 346 & 25 & 724 & 48.76 \\
 & Gemini-3.1-Pro & 372 & 342 & 10 & 724 & 51.38 \\
\cmidrule[0.5pt](lr){2-7}
 & DeepSeek-V3.2 & 221 & 265 & 238 & 724 & 30.52 \\
 & Kimi-K2.5 & 307 & 396 & 21 & 724 & 42.40 \\
 & Qwen3.5-397B & 358 & 362 & 4 & 724 & 49.45 \\
 & Qwen3-235B & 271 & 299 & 154 & 724 & 37.43 \\
 & Qwen3-30B & 220 & 323 & 181 & 724 & 30.39 \\
 & MiniMax-M2.5 & 316 & 390 & 18 & 724 & 43.65 \\
 & Llama-4-Scout & 234 & 392 & 98 & 724 & 32.32 \\
\midrule
\multirow{10}{*}{\rotatebox{90}{\textbf{Inline}}} & GPT-5.4 & 352 & 351 & 21 & 724 & 48.62 \\
 & Claude-Sonnet-4.6 & 53 & 60 & 611 & 724 & 7.32 \\
 & Gemini-3.1-Pro & 132 & 146 & 446 & 724 & 18.23 \\
\cmidrule[0.5pt](lr){2-7}
 & DeepSeek-V3.2 & 113 & 129 & 482 & 724 & 15.61 \\
 & Kimi-K2.5 & 98 & 148 & 478 & 724 & 13.54 \\
 & Qwen3.5-397B & 71 & 79 & 574 & 724 & 9.81 \\
 & Qwen3-235B & 50 & 68 & 606 & 724 & 6.91 \\
 & Qwen3-30B & 117 & 183 & 424 & 724 & 16.16 \\
 & MiniMax-M2.5 & 47 & 73 & 604 & 724 & 6.49 \\
 & Llama-4-Scout & 59 & 114 & 551 & 724 & 8.15 \\
\bottomrule
\end{tabular}
\end{table*}

\textbf{Dynamic pass rates confirm and sharpen the static picture: inline mode fails predominantly at the installation stage rather than at the logic stage.}
Table~\ref{tab:pipeline_d1_bcb_results_osv_subset} reports BigCodeBench execution outcomes.
Under explicit prompting, pass rates range from 30.39\% to 51.38\%, broadly consistent with the static compatibility rates in §\ref{sec:rq3_static}.
Under inline prompting, pass rates collapse to 6.49\%--48.62\%, with nine of ten models below 20\%.
The critical diagnostic is the \#Error column, which captures tasks that terminate before any test assertion can run, typically due to installation failure.
Under inline mode, error counts reach 424--611 per model across the 724-task benchmark, meaning that the majority of tasks never execute at all.
Under explicit prompting, error counts are overwhelmingly lower, reaching single digits for most closed-source models, which confirms that the inline-explicit gap is attributable to version-level incompatibility rather than any difference in code generation quality between the two modes.

GPT-5.4 is the notable exception: its inline pass rate of 48.62\% nearly matches its explicit rate of 47.93\%, with only 21 error-category tasks.
This is consistent with its static compatibility advantage observed in §\ref{sec:rq3_static}, where it was the only model achieving above 60\% in both modes.

\subsubsection{Relationship Between Static and Dynamic Compatibility}
\label{sec:rq3_sankey}

\begin{figure}[t]
  \centering
  \includegraphics[width=\linewidth]{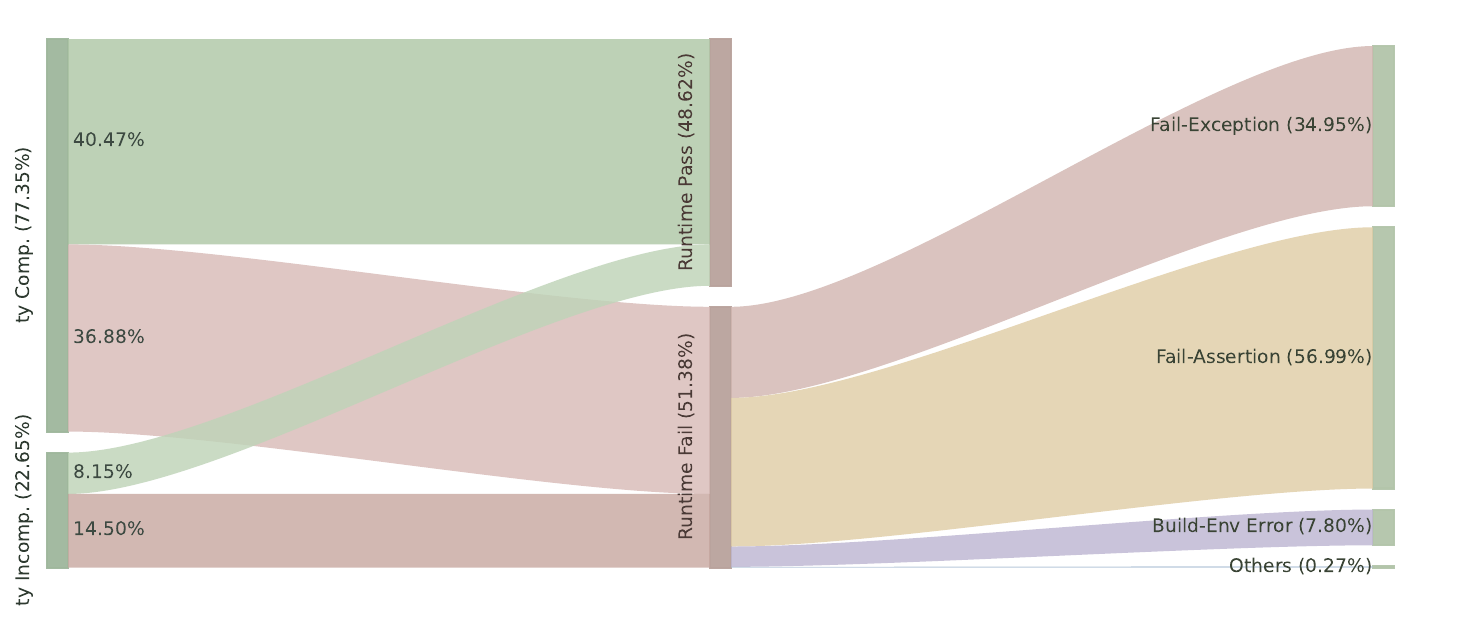}
  \Description{A Sankey diagram showing the flow from static compatibility outcomes to dynamic execution results across 724 BigCodeBench tasks under inline prompting mode for GPT-5.4. Left nodes represent the ty static type checker verdict (compatible or incompatible); right nodes represent BigCodeBench runtime outcomes (Pass, Fail-Assertion, Fail-Exception, Build/Env Error, and Others). Proportions are relative to all 724 tasks. The diagram shows that 77.35 percent of tasks are classified as ty-compatible, of which 40.47 percent achieve a runtime pass, while 36.88 percent fail at runtime despite passing static checking. Among ty-incompatible tasks, 8.15 percent nonetheless pass at runtime, indicating a residual rate of false positive static errors.}
  \caption{Flow from static compatibility outcomes to dynamic execution results
    across BigCodeBench tasks under inline mode, shown for GPT-5.4.
    Left nodes represent \texttt{ty} static verdict, right nodes represent BCB runtime outcome.
    Proportions are relative to all 724 tasks in the first and second columns. The proportions in the last column are relative to the 372 runtime fail tasks.}
  \label{fig:sankey}
\end{figure}

\textbf{Static compatibility is a strong positive predictor of dynamic success, but the two measures are not interchangeable.}
Figure~\ref{fig:sankey} illustrates the relationship for GPT-5.4 under inline mode, the model with the highest inline pass rate and therefore the most interpretable signal.
Of all 724 tasks, 77.35\% are classified as static compatible and 22.65\% as incompatible.
Among the compatible tasks, 40.47\% flow through to a runtime pass, while 36.88\% fail at runtime despite passing static checking.
Within the compatible set, runtime failures decompose into four categories.
Fail-Assertion accounts for the largest share at 56.99\%, representing tasks where the API surface is correct but the generated code is functionally wrong.
Fail-Exception covers uncaught runtime exceptions at 34.95\%.
Build/Env Error captures environment setup failures that occur after static checking passes, at 7.80\%.
The remaining 0.27\% falls into other categories.
All of these errors that static type checking cannot detect in advance.

In the reverse direction, 8.15\% of tasks classified as \texttt{ty} static incompatible nonetheless pass at runtime, indicating that a fraction of static type errors are false positives or reflect API differences that do not affect actual execution paths.
Static compatibility therefore substantially overestimates the runtime pass rate when the incompatible fraction is large, as is the case for most models under inline mode.
It does, however, provide a reliable lower bound on the fraction of tasks that will fail.

\begin{summary}
\textbf{Summary:} Static compatibility rates range from 19.70\% to 63.20\% under inline prompting and 45.30\%--66.70\% under explicit prompting, with dependency installation failure due to temporal lag identified in RQ1 as the dominant cause.
Specifically, LLMs preferentially recommend older versions that disproportionately fail to install in modern Python environments.
Dynamic verification confirms the pattern, with inline pass rates collapsing to 6.49\%--48.62\% as version incompatibilities prevent most tasks from executing at all.
\end{summary}


\section{Robustness, Diagnosis, and Mitigation}
\label{sec:rdm}

The main findings in §\ref{sec:results} raise three follow-up questions: Do the compatibility results hold across different Python runtime environments? Are the installation failures caused by version selection or by the generated code itself? Can prompt-level interventions reduce the identified risks?
§\ref{sec:robustness} addresses the first question by repeating the full compatibility pipeline across four Python versions.
§\ref{sec:diagnosis} addresses the second through a controlled substitution experiment that holds generated code fixed and varies only the specified versions.
§\ref{sec:mitigation} addresses the third by evaluating four progressively stronger prompt-level interventions.

\subsection{Robustness Across Python Environments}
\label{sec:robustness}

The main experiments use Python 3.12 as the execution environment.
To assess whether the compatibility findings generalize across runtime versions, we repeat the full Stage (5) pipeline under Python 3.8, 3.10, and 3.14 on the BigCodeBench dataset.
This dataset supports both static and dynamic evaluation on the same task set.
Vulnerability exposure is not re-evaluated. It is a static property of $(\ell,v)$ pairs and is invariant to the runtime environment.

Table~\ref{tab:pipeline_d1_compat_by_python} reports static compatibility and dynamic pass rates across all four Python versions and both prompting modes.

\begin{table*}[t]\small
\centering
\caption{Static compatibility ($\tau_C$, \%) and BigCodeBench dynamic pass rate (\%) on the BigCodeBench dataset across Python runtime versions.}
\label{tab:pipeline_d1_compat_by_python}
\begin{tabular}{llrrrrrrrr}
\toprule
& \textbf{Model} & \multicolumn{2}{c}{\textbf{3.8}} & \multicolumn{2}{c}{\textbf{3.10}} & \multicolumn{2}{c}{\textbf{3.12}} & \multicolumn{2}{c}{\textbf{3.14}} \\
\cmidrule(lr){3-4}\cmidrule(lr){5-6}\cmidrule(lr){7-8}\cmidrule(lr){9-10}
& & $\tau_C$ & BCB & $\tau_C$ & BCB & $\tau_C$ & BCB & $\tau_C$ & BCB \\
\midrule
\multirow{10}{*}{\rotatebox{90}{\textbf{Explicit}}} & GPT-5.4 & 2.95 & 1.72 & 14.76 & 8.12 & 91.64 & 49.32 & 91.14 & 49.08 \\
 & Claude-Sonnet-4.6 & 8.00 & 5.04 & 16.24 & 8.49 & 88.81 & 49.32 & 88.56 & 49.20 \\
 & Gemini-3.1-Pro & 3.94 & 2.21 & 15.99 & 8.12 & 92.74 & 52.03 & 92.25 & 51.54 \\
\cmidrule[0.5pt](lr){2-10}
 & DeepSeek-V3.2 & 44.16 & 23.49 & 56.58 & 27.43 & 63.84 & 31.86 & 54.98 & 28.17 \\
 & Kimi-K2.5 & 81.67 & 39.85 & 74.54 & 40.71 & 88.56 & 43.42 & 87.33 & 42.80 \\
 & Qwen3.5-397B & 4.67 & 2.83 & 15.74 & 8.86 & 92.74 & 50.55 & 92.50 & 50.31 \\
 & Qwen3-235B & 44.90 & 22.02 & 45.51 & 23.25 & 74.42 & 38.50 & 74.54 & 38.50 \\
 & Qwen3-30B & 51.91 & 27.18 & 52.15 & 27.80 & 66.54 & 31.73 & 65.44 & 31.12 \\
 & MiniMax-M2.5 & 10.46 & 5.41 & 19.31 & 9.84 & 91.14 & 44.53 & 91.14 & 44.28 \\
 & Llama-4-Scout & 6.40 & 3.57 & 16.61 & 7.26 & 80.81 & 33.21 & 80.69 & 32.96 \\
\midrule
\multirow{10}{*}{\rotatebox{90}{\textbf{Inline}}} & GPT-5.4 & 9.47 & 4.92 & 74.78 & 46.86 & 78.72 & 49.69 & 27.80 & 12.55 \\
 & Claude-Sonnet-4.6 & 75.28 & 42.44 & 87.95 & 30.14 & 21.89 & 9.59 & 17.34 & 9.10 \\
 & Gemini-3.1-Pro & 63.35 & 35.30 & 92.25 & 35.67 & 41.08 & 21.03 & 19.19 & 8.36 \\
\cmidrule[0.5pt](lr){2-10}
 & DeepSeek-V3.2 & 57.56 & 28.04 & 90.77 & 38.25 & 34.19 & 16.73 & 13.90 & 5.66 \\
 & Kimi-K2.5 & 59.78 & 31.12 & 90.65 & 32.47 & 37.27 & 14.15 & 15.62 & 6.52 \\
 & Qwen3.5-397B & 81.43 & 45.26 & 93.60 & 28.78 & 25.95 & 10.95 & 18.20 & 8.49 \\
 & Qwen3-235B & 62.73 & 29.52 & 66.79 & 17.59 & 32.84 & 8.24 & 22.51 & 9.72 \\
 & Qwen3-30B & 48.59 & 24.11 & 62.12 & 22.14 & 45.51 & 17.84 & 36.04 & 15.74 \\
 & MiniMax-M2.5 & 79.83 & 40.47 & 91.88 & 30.75 & 21.28 & 8.12 & 15.74 & 6.77 \\
 & Llama-4-Scout & 61.99 & 25.95 & 68.02 & 19.19 & 26.20 & 9.10 & 25.95 & 9.23 \\
\bottomrule
\end{tabular}
\end{table*}

\textbf{The two prompting modes respond to Python version in opposite directions, revealing that the compatibility pattern is driven by LLM version choices rather than by the runtime itself.}
Under inline mode, both static compatibility and dynamic pass rates peak around Python 3.10 and decline sharply at 3.12 and beyond.
LLMs preferentially specify library versions released before 2022, and many of those versions carry build-time dependencies on \texttt{distutils}, which was available through Python 3.10 but removed from the standard library in Python 3.12.
For DeepSeek-V3.2, inline $\tau_C$ falls from 90.77\% at Python 3.10 to 34.19\% at 3.12.
The BCB pass rate drops from 38.25\% to 16.73\% over the same transition.
Under explicit mode, both metrics increase monotonically from Python 3.8 to 3.12 and then stabilize, because LLM outputs of most explicit mode tasks rely on the latest version fallback, and newer releases have progressively dropped support for older versions of Python.
The reversal between the two mode trends confirms that the sensitivity to Python version is entirely a consequence of version choices made by the LLM, not a property of the runtime.

\textbf{Static and dynamic results track each other closely across all Python versions.}
For every model and every Python version, the BCB dynamic pass rate moves in the same direction as $\tau_C$.
This confirms that installation failure is the governing mechanism: a version that cannot install prevents execution before any test can run.

\textbf{Cross-model rankings are preserved across Python environments.}
The ordering of models by compatibility under inline mode is consistent across all four Python versions, matching the ordering reported in §\ref{sec:rq3}.
The choice of Python 3.12 as the primary evaluation environment does not introduce systematic bias in cross-model comparisons.

\subsection{Diagnosing Compatibility Failures}
\label{sec:diagnosis}

Section~\ref{sec:rq3} showed that installation failure is the dominant cause of static incompatibility under inline prompting.
The remaining question is whether that failure stems from version selection or from the generated code itself imposing unsatisfiable constraints.
The two causes require different remedies, so a controlled attribution is necessary.

\textbf{Experimental design.}
To isolate version selection as the cause of installation failures, we hold the generated code fixed and test whether swapping the specified version to a neighboring release resolves the failure.
We select qualifying tasks from the inline-mode \myso results under two conditions: the task must have failed Stage (5) by dependency installation error, and all LLM-specified versions in the task must be CVE-free.
The first condition ensures the failure is unambiguously at the dependency resolution layer.
The second eliminates security as a confounding variable: if a vulnerable version were included, a neighboring version might resolve the failure by switching to a safer release, entangling compatibility and security attribution.

For each qualifying task, we hold the generated code $G(t,m,p)$ fixed and search for a substitute version.
Let $\mathcal{V}_{\mathrm{fail}}(t,m,p) \subseteq \mathcal{V}(t,m,p)$ denote the subset of LLM-specified, version-annotated library with install error.
For each $(\ell, v) \in \mathcal{V}_{\mathrm{fail}}$, we enumerate a neighbor version as:
\begin{equation}
    \mathcal{N}(\ell, v) = \{\, v' \in \Pi(\ell) : |r(v') - r(v)| \leq N \,\}
\end{equation}
where $\Pi(\ell)$ is the ordered list of published PyPI releases for library $\ell$ and $r(\cdot)$ maps a version to its release-sequence index.
We set $N = 3$, yielding up to six neighbors per library, a window narrow enough to avoid crossing major API boundaries and introducing breaking changes between versions.
Candidate substitutions are formed by replacing the failing version in $\mathcal{V}_{\mathrm{fail}}$ with a neighbor while holding all other versions fixed.
To bound the search, we cap the number of evaluated combinations per task at $k = 40$.
For each evaluated substitution, we re-run Stage (5). A task is \emph{installable-recovered} if the substitution removes all installation error occurrences, and \emph{compatible-recovered} if it additionally passes \texttt{ty} static type checking.
All reported rates are lower bounds, since the neighborhood is finite and may not cover all compatible releases.

\textbf{Version substitution resolves installation failures in 17.19\%--55.00\% of qualifying tasks, establishing version selection as the primary cause.}
Table~\ref{tab:neighbor_ty_install_error_recovery_d2_py312} reports results for all ten models.
The compatible rate, the more demanding criterion requiring both installation success and static compatibility, ranges from 7.81\% to 46.04\%.
Recovery rates vary across models, reflecting differences in how badly each model's version choices deviate from installable releases.
The phenomenon is consistent, however: in every case, a neighboring version resolves failures that the originally specified version cannot.
Since the generated code is held fixed throughout, any resolution is attributable solely to the version change.

\textbf{Neighboring-version substitution introduces negligible additional security risk.}
Across all 64--205 qualifying tasks per model, only five total recovered tasks across all models introduce at least one new CVE in the substitute version.
Six models report zero such cases, while four models report one or two each.
Compatibility recovery and security safety are therefore largely compatible objectives: the version neighborhood that resolves installation failures is not, in general, a more vulnerable region of the release timeline.

\textbf{When installation succeeds, static type checking also passes in the majority of cases.}
Among tasks where a neighboring version removes the installation error, the fraction that additionally passes the static type check is consistently high.
For example, Gemini-3.1-Pro recovers 91 tasks at the installable level, of which 76 (83.5\%) also pass static type checking.
DeepSeek-V3.2 recovers 95, of which 78 (82.1\%) pass.
The high conditional type-check rate confirms that the generated code is largely API-consistent once the version constraint is resolved. Thus, the original failure was caused by the version choice, not by a defect in the generated logic.

\begin{table*}[t]\small
\centering
\caption{Neighbor-version search on \myso tasks that have package installation errors.
  \textbf{Qualifying} is the number of tasks meeting the selection criteria.
  \textbf{Installable} counts tasks with a neighbor that removes the installation error (percentage of Qualifying).
  \textbf{Compatible} counts tasks with a neighbor that additionally passes static type-checking (percentage of Qualifying).
  \textbf{Vuln@Reco} counts installable recoveries that introduce at least one new CVE.
  $\overline{N}_{\mathrm{test}}$ is the mean number of neighbor candidates evaluated per qualifying task.
  All rates are lower bounds under the configured search budget.}
\label{tab:neighbor_ty_install_error_recovery_d2_py312}
\begin{tabular}{lrrrrr}
\toprule
\textbf{Model} & \textbf{Qualifying} & \textbf{Installable} & \textbf{Compatible} & \textbf{Vuln@Reco} & $\mathbf{\overline{N}_{\mathrm{test}}}$ \\
\midrule
GPT-5.4           &  64 &  11 (17.19\%) &   5 (7.81\%)  & 1 & 14.80 \\
Claude-Sonnet-4.6 &  92 &  28 (30.43\%) &  15 (16.30\%) & 0 & 15.12 \\
Gemini-3.1-Pro    & 177 &  91 (51.41\%) &  76 (42.94\%) & 0 & 14.58 \\
\cmidrule[0.5pt](lr){1-6}
DeepSeek-V3.2     & 202 &  95 (47.03\%) &  78 (38.61\%) & 0 & 16.24 \\
Kimi-K2.5         & 173 &  78 (45.09\%) &  60 (34.68\%) & 0 & 18.10 \\
Qwen3.5-397B      & 202 & 109 (53.96\%) &  93 (46.04\%) & 0 & 15.70 \\
Qwen3-235B        & 205 & 104 (50.73\%) &  80 (39.02\%) & 0 & 18.43 \\
Qwen3-30B         & 140 &  77 (55.00\%) &  57 (40.71\%) & 1 & 15.16 \\
MiniMax-M2.5      & 197 & 103 (52.28\%) &  28 (14.21\%) & 2 & 20.78 \\
Llama-4-Scout     & 205 &  50 (24.39\%) &  34 (16.59\%) & 1 & 21.20 \\
\bottomrule
\end{tabular}
\end{table*}

\subsection{Mitigation Probe}
\label{sec:mitigation}

Section~\ref{sec:diagnosis} established that version selection is the primary cause of compatibility failures, and §\ref{sec:rq2} showed that LLM-specified versions frequently carry known CVEs.
We evaluate whether these problems can be reduced through prompt-level interventions, without modifying model weights.

\textbf{Experimental conditions.}
\textbf{Baseline} (\texttt{inline}) is the standard inline prompt from §\ref{sec:results}.
\textbf{Safety instruction} (\texttt{abl-instruct}) appends a natural-language instruction to avoid versions with known CVEs.
No structured version information is provided.
\textbf{Version anchor} (\texttt{abl-version}) injects a pre-computed safe version for each TPL determined by the OSV index and PyPI metadata, following a three-tier priority: the most recent CVE-free release, the release with the fewest known CVEs, or the most recent known release as a last resort.
Specifically, 91.33\% of TPLs in our dataset were resolved at the safe tier, 8.67\% at min-vuln.
\textbf{Version anchor with RAG} (\texttt{abl-rag}) extends \texttt{abl-version} by additionally retrieving the top-20 most relevant API signatures for each resolved $(\ell,v)$ pair.
API signatures for each resolved $(\ell,v)$ pair are first extracted from the installed package using \texttt{griffe}~\cite{griffe}, a static analysis tool that parses Python source and stubs to produce structured API documentation. The top-20 most relevant signatures are then selected by BM25 retrieval against the task question text and prepended to the prompt.


\begin{table*}[t]\small
\centering
\caption{Security metrics across ablation conditions on the \myso dataset.
  $\rho_U(\%)$: library vulnerability rate. $\tau_U(\%)$: task vulnerability exposure.}
\label{tab:ablation_security}
\begin{tabular}{lrrrrrrrr}
\toprule
 & \multicolumn{2}{c}{\textbf{Baseline}} & \multicolumn{2}{c}{\textbf{abl-instruct}} & \multicolumn{2}{c}{\textbf{abl-version}} & \multicolumn{2}{c}{\textbf{abl-rag}} \\
\cmidrule(lr){2-3}\cmidrule(lr){4-5}\cmidrule(lr){6-7}\cmidrule(lr){8-9}
\textbf{Model} & $\rho_U$ & $\tau_U$ & $\rho_U$ & $\tau_U$ & $\rho_U$ & $\tau_U$ & $\rho_U$ & $\tau_U$ \\
\midrule
GPT-5.4           & 40.51 & 46.20 & 38.39 & 44.00 & 9.98  & 10.90 & 10.47 & 11.60 \\
Claude-Sonnet-4.6 & 38.58 & 50.50 & 46.55 & 54.80 & 11.99 & 15.00 & 14.88 & 16.00 \\
Gemini-3.1-Pro    & 42.13 & 50.70 & 42.76 & 51.90 & 12.83 & 15.80 & 11.55 & 13.80 \\
\cmidrule{1-9}
DeepSeek-V3.2     & 40.76 & 54.00 & 40.90 & 54.60 & 19.34 & 25.00 & 17.91 & 21.30 \\
Kimi-K2.5         & 45.78 & 55.70 & 44.61 & 54.90 & 20.83 & 22.30 & 20.06 & 21.80 \\
Qwen3.5-397B      & 40.35 & 50.40 & 40.39 & 49.40 & 12.13 & 15.80 & 13.30 & 16.40 \\
Qwen3-235B        & 37.60 & 51.70 & 36.40 & 50.80 & 15.62 & 17.00 & 16.55 & 19.30 \\
Qwen3-30B         & 32.85 & 36.70 & 32.23 & 36.80 & 22.17 & 23.00 & 24.00 & 24.20 \\
MiniMax-M2.5      & 45.42 & 52.10 & 45.05 & 50.20 & 15.02 & 19.10 & 14.87 & 17.70 \\
Llama-4-Scout     & 38.59 & 51.70 & 38.36 & 49.90 & 21.08 & 27.60 & 20.53 & 26.60 \\
\bottomrule
\end{tabular}
\end{table*}

\textbf{Safety instruction provides negligible security improvement.}
As shown in Table~\ref{tab:ablation_security}, across all ten models, \texttt{abl-instruct} reduces $\rho_U$ by at most 2.2 percentage points and $\tau_U$ by at most 2.2 percentage points relative to baseline.
For three models whose baseline exposure is already partially grounded in version awareness, Gemini-3.1-Pro, DeepSeek-V3.2, and Qwen3.5-397B, $\rho_U$ and $\tau_U$ are essentially unchanged or slightly higher under \texttt{abl-instruct}.
The negligible improvement is consistent with the temporal analysis in §\ref{sec:rq2_temporal}: the relevant CVEs were publicly disclosed before each model's training cutoff, so the model is not unaware of the vulnerability landscape in principle.
Rather, it lacks the ability to map version identifiers to CVE records at inference time, and a natural-language instruction cannot compensate for this grounding gap.

\textbf{Version anchoring substantially reduces but does not eliminate vulnerability exposure.}
Under \texttt{abl-version}, library vulnerability rate $\rho_U$ drops to 9.98\%--22.17\% and $\tau_U$ to 10.90\%--27.60\% across models, as shown in Table~\ref{tab:ablation_security}.
These are large reductions relative to baseline, but residual exposure remains.
The residual arises when no CVE-free release exists for a given library: the version anchor falls back to the release with the fewest known CVEs rather than a CVE-free one.
For such libraries, any pinned version carries at least one CVE, making zero exposure structurally unachievable without excluding the library entirely.
However, all ten models report $\rho_U$ values under \texttt{abl-version} that exceed this 8.67\% floor of the \myso dataset itself, indicating that the models introduce additional TPLs beyond those specified in the injected version list, and some of those self-introduced libraries are themselves pinned to vulnerable versions.
\texttt{abl-rag} yields similar security metrics, confirming that the residual vulnerability is a property of the library coverage rather than a consequence of the API-grounding intervention.

\textbf{Version anchoring systematically improves compatibility.}
The compatibility effect of the ablation \texttt{abl-version} is far larger than its security effect, as Table~\ref{tab:ablation_compat} shows.
On BigCodeBench, $\tau_C$ increases from a baseline range of 21.89\%--78.72\% to 80.92\%--93.64\% under \texttt{abl-version}, and dynamic pass rates rise from 7.32\%--49.69\% to 36.31\%--54.56\%.
The improvement is uniform across model families, indicating that the baseline compatibility failures are attributable to version selection rather than to model-level code generation quality.

\textbf{RAG-augmented prompting provides marginal additional benefit.}
\texttt{abl-rag} improves $\tau_C$ by 0.10--2.05 percentage points over \texttt{abl-version} and dynamic pass rates by at most 1.03 percentage points, as the $\Delta$ columns in Table~\ref{tab:ablation_compat} show.
For some models, the dynamic pass rate under \texttt{abl-rag} is slightly lower than under \texttt{abl-version}, suggesting that injected API context occasionally misleads rather than assists.
Two structural factors limit the approach.
First, BM25 retrieval uses the task question text as the query, which describes the programming goal rather than the specific functions the model will invoke; retrieved signatures may therefore cover API surface adjacent to, but not identical to, what the generated code actually needs.
Second, injected signatures for one version may conflict with the model's parametric knowledge of a different version's API, introducing inconsistency that the model cannot resolve at inference time.
The marginal and occasionally negative gain is consistent with both factors: when retrieval is well-matched, it provides modest benefit; when it is mismatched, it can degrade generation.
The residual compatibility failures are therefore more likely attributable to limitations in version-API co-generation ability~\cite{liu2025codeupdatearena,11224651,11121691,10.1145/3808230} than to ignorance of the API surface, which needs further design that is outside this paper's scope.

\begin{table*}[t]\small
\centering
\caption{Static compatibility ($\tau_C$, \%) and BigCodeBench dynamic pass rate (\%) across ablation conditions.
Values in parentheses give percentage-point change versus Baseline.}
\label{tab:ablation_compat}
\resizebox{\linewidth}{!}{
\begin{tabular}{lrr rr rr rr}
\toprule
 & \multicolumn{2}{c}{\textbf{Baseline}}
 & \multicolumn{2}{c}{\textbf{abl-instruct}}
 & \multicolumn{2}{c}{\textbf{abl-version}}
 & \multicolumn{2}{c}{\textbf{abl-rag}} \\
\cmidrule(lr){2-3}\cmidrule(lr){4-5}\cmidrule(lr){6-7}\cmidrule(lr){8-9}
\textbf{Model}
  & $\tau_C$ & BCB
  & $\tau_C$ & BCB
  & $\tau_C$ & BCB
  & $\tau_C$ & BCB \\
\midrule
GPT-5.4 & 77.35 & 48.62 & $77.35(+0.00)$ & $46.82(-1.80)$ & $91.27(+13.92)$ & $46.67(-1.95)$ & $91.75(+14.40)$ & $46.19(-2.43)$ \\
Claude-Sonnet-4.6 & 16.16 & 7.32 & $37.98(+21.82)$ & $23.62(+16.30)$ & $90.48(+74.32)$ & $45.87(+38.55)$ & $89.37(+73.20)$ & $41.90(+34.58)$ \\
Gemini-3.1-Pro & 35.91 & 18.23 & $42.13(+6.22)$ & $23.62(+5.39)$ & $92.22(+56.31)$ & $52.22(+33.99)$ & $93.17(+57.26)$ & $52.38(+34.15)$ \\
\cmidrule{1-9}
DeepSeek-V3.2 & 31.63 & 15.61 & $38.12(+6.49)$ & $18.92(+3.31)$ & $88.25(+56.62)$ & $41.27(+25.66)$ & $90.00(+58.37)$ & $42.38(+26.77)$ \\
Kimi-K2.5 & 32.60 & 13.54 & $60.22(+27.62)$ & $28.73(+15.19)$ & $89.84(+57.24)$ & $42.38(+28.85)$ & $90.16(+57.56)$ & $43.49(+29.96)$ \\
Qwen3.5-397B & 20.03 & 9.81 & $37.29(+17.27)$ & $20.58(+10.77)$ & $89.68(+69.65)$ & $46.83(+37.02)$ & $90.16(+70.13)$ & $45.87(+36.07)$ \\
Qwen3-235B & 28.73 & 6.91 & $26.93(-1.80)$ & $7.46(+0.55)$ & $87.14(+58.41)$ & $41.90(+35.00)$ & $89.52(+60.79)$ & $43.81(+36.90)$ \\
Qwen3-30B & 41.57 & 16.16 & $43.37(+1.80)$ & $15.75(-0.41)$ & $86.83(+45.25)$ & $41.11(+24.95)$ & $86.67(+45.09)$ & $40.32(+24.16)$ \\
MiniMax-M2.5 & 15.88 & 6.49 & $28.31(+12.43)$ & $13.54(+7.04)$ & $88.89(+73.00)$ & $45.08(+38.59)$ & $88.89(+73.00)$ & $43.65(+37.16)$ \\
Llama-4-Scout & 23.34 & 8.15 & $20.99(-2.35)$ & $7.87(-0.28)$ & $78.57(+55.23)$ & $32.38(+24.23)$ & $80.63(+57.29)$ & $31.90(+23.76)$ \\
\bottomrule
\end{tabular}
}
\end{table*}


\section{Discussion}
\label{sec:discussion}

LLMs routinely introduce version annotations that are vulnerable, incompatible, or both, driven by a shared bias rather than individual model defects.
The problem resists prompt-level remediation but is addressable through external tooling.
This section draws practical implications for developers, LLM providers, and the research community.

\subsection{The Problem Is Ecosystem-Level, Not Model-Level}
\label{sec:discussion_systemic}

The cross-model convergence documented in §\ref{sec:rq2_convergence} is the most structurally significant finding of this study.
\texttt{django==6.0.1}, \texttt{requests==2.31.0}, and \texttt{flask==3.1.2} appear among the most-specified vulnerable pairs for at least nine of the ten evaluated models, with counts from closed-source and open-source families alike confirming the pattern as illustrated in Figure~\ref{fig:vuln_convergence}.
No individual model accounts for this pattern, which reflects a shared distribution in the training corpora.

The mechanism follows directly from the temporal lag established in §\ref{sec:rq1_temporal}: versions released years before a model's knowledge cutoff accumulate far greater training signal than recent ones, and those high-salience versions are precisely the ones that have had time to accumulate unpatched CVEs.
The median lag ranges from 9 to 31 months across all ten models.
This salience-vulnerability coupling is a structural consequence of web-crawled training data, not a correctable defect in any individual model, which is why practitioners cannot reduce exposure by switching models.
Risk must be addressed at the point of version resolution.

\subsection{Implications for Developers}
\label{sec:discussion_developers}

\textbf{LLM-specified versions require independent security validation before adoption.}
As documented in §\ref{sec:rq2_overall}, over a third to more than half of all inline-prompting tasks contain a version with a known CVE, and the majority of those vulnerable versions are rated \textit{Critical} or \textit{High}.
Accepting LLM-generated version annotations without review is sampling from a distribution heavily skewed toward high-severity vulnerabilities.
Standard practice should route LLM-specified versions through an SCA tool or OSV query before committing them to a dependency manifest.

\textbf{The prompting mode is itself a risk variable, and neither mode is universally safe.}
Switching from inline to explicit mode reduces the version specification rate by up to 88.73 percentage points as in §\ref{sec:rq1_spec}.
This substantially lowers the surface area for version-level vulnerability, since unspecified dependencies cannot carry known CVEs.
However, from the results in Table~\ref{tab:pipeline_d1_compat_by_python}, explicit mode introduces a different risk: those unspecified dependencies are resolved to the latest available release at install time, which collapses compatibility under older Python environments for most models.
In production codebases, explicit mode needs to be carefully used because it may omit dependency versions. In exploratory contexts, inline mode output should be treated as an audit signal rather than a trusted declaration. Overall, developers should include package security audit tools in the coding pipeline \textit{e.g.}, \texttt{pip-audit}~\cite{pip-audit}, to improve software supply chain security in the LLM-assisted coding.

\textbf{Minor version substitution improves compatibility but not security.}
As discussed in §\ref{sec:rq2_safe}, 95.11\% of vulnerable $(\ell,v)$ pairs have no CVE-free release in the same major version branch, meaning that a patch-level increment is structurally insufficient to escape vulnerability exposure.
On the compatibility dimension, the diagnosis experiment in §\ref{sec:diagnosis} shows that neighboring-version search does recover installation failures in 17\%--55\% of cases, but these recovered versions are not guaranteed to be CVE-free.
Developers should therefore anticipate a cross-major migration or a security-aware resolver rather than a simple patch-level increment.

\subsection{Implications for LLM Providers}
\label{sec:discussion_providers}

\textbf{The version-selection problem is a grounding gap, not a safety alignment gap, and cannot be closed through prompting alone.}
As shown in Table~\ref{tab:ablation_security}, the \texttt{abl-instruct} condition reduces $\rho_U$ by at most 2.2 percentage points across all ten models.
The CVE disclosure analysis in §\ref{sec:rq2_temporal} explains why: 72\%--91\% of the CVEs carried by LLM-specified versions were already publicly disclosed before each model's training cutoff, so the model is not unaware of the vulnerability landscape.
The failure is in mapping a version identifier to its CVE status at generation time.
Addressing this requires changes to the training data or the inference architecture, not to the prompt.
On the training side, the version-salience bias in §\ref{sec:rq1_temporal} points to a concrete intervention: flagging $(\ell,v)$ pairs that appear frequently in training text while carrying an active OSV record, then de-weighting or annotating them with a security signal.
The fact that \texttt{abl-version} reduces $\rho_U$ by 20--36 percentage points relative to baseline confirms that models can follow externally provided version constraints; the gap is in spontaneous selection.
 
\textbf{Inference-time information injecting can help.}
No training-time intervention can cover vulnerability disclosures that post-date the cutoff. The 9--31 month median lag will persist as long as models rely on static snapshots.
Architectures that query a live vulnerability feed at inference time address this limitation directly.
Separately, current code generation benchmarks measure logical correctness and functional pass rates but do not evaluate whether the dependency versions a model selects are safe, valid, or installable.
The measurement framework developed in this study provides a reusable template for such evaluation, and incorporating version-annotation quality into routine benchmarks would create direct incentives for providers to close the grounding gap.

\textbf{Community disclosure has produced initial acknowledgment from model providers.}
We disclosed the observed version-selection behavior to the community of all 10 evaluated models and to AI coding assistant providers, including GitHub Copilot and Cursor.
Several parties have responded.
OpenAI's support channel acknowledged that \textit{``right now there's no built-in CVE check when suggesting versions, so this gap shows up across models...''}\footnote{\url{https://community.openai.com/t/1378999}} and indicated that the feedback would be forwarded to the product team.
Alibaba Cloud and Moonshot AI similarly confirmed receipt of the findings through their respective support channels.
These responses corroborate that the version-selection gap reflects a recognized but currently unaddressed behavior in LLMs.

\section{Threats to Validity}
\label{sec:threats}

In this section, we report the internal and external validity problems of this study and discuss the measures we take to relieve their impact.
 
\subsection{Internal Validity}
 
\textbf{Construct operationalization.}
Static compatibility is operationalized through installation success and \texttt{ty} static type checking, neither of which is equivalent to full runtime correctness.
\texttt{ty} does not detect semantic errors in generated logic or behaviors not captured by type stubs.
We mitigate this by complementing Stage (5) with BigCodeBench dynamic execution as designed in §\ref{sec:rq3_dynamic}. 
The Sankey analysis in §\ref{sec:rq3_sankey} quantifies the residual gap between the two signals.
Vulnerability exposure is operationalized as CVE presence in the OSV index, which is a conservative upper bound on exploitability rather than a precise risk measure.
Exploitability assessment would require per-task taint analysis, which is beyond this study's scope.

\textbf{Static type checker bias.}
$\tau_C$ depends on \texttt{ty}, which may both under- and over-report incompatibility.
The Sankey analysis in §\ref{sec:rq3_sankey} bounds the false positive rate at 8.15\%: that fraction of \texttt{ty}-incompatible tasks still pass BigCodeBench execution.
Libraries lacking type stubs are largely invisible to \texttt{ty}, causing $\tau_C$ to overestimate compatibility for poorly-typed packages.
The BigCodeBench dynamic evaluation mitigates both concerns, as execution evidence is not subject to stub coverage constraints.

 
\textbf{Sampling variance.}
We generate one output per $(t, m, p)$ triple using provider-default temperature settings, reflecting the deployment conditions most developers encounter.
Characterizing within-model variance in version annotation is outside the scope of this study. Future work could sample multiple outputs per triple to bound this variability.
 
\textbf{Prompting mode coverage.}
The two evaluated modes cover the most common developer interaction patterns.
Agentic workflows, multi-turn refinement, and IDE-level system prompts may exhibit different version-annotation behavior, and we make no claims about these settings.
Our study establishes a baseline for the most frequent interaction modes. Thus, characterizing the full space of prompting contexts is a direction for future work.

\subsection{External Validity}
 
\textbf{Ecosystem scope.}
This study is scoped to Python and PyPI.
Version-pinning risks plausibly exist in other ecosystems, \textit{e.g.}, npm, Maven, Cargo, but vulnerability exposure rates, compatibility failure patterns, and model behavior may differ substantially.
We make no generalizability claims beyond Python, and it can be a good direction for future work.
 
\textbf{Dataset representativeness.}
The \myso dataset employs TPL-balanced sampling, but the underlying question distribution still favors high-activity libraries and pre-LLM-adoption periods, \textit{i.e.}, 313 tasks from 2020 versus 40 from 2025.
Niche libraries and post-2024 development contexts may exhibit different version-annotation behavior than what we observe. This is a limitation caused by the huge reduction to 10\% of remaining questions on Stack Overflow itself~\cite{so_query} as discussed in section \S\ref{sec:setup_dataset}, which may need new dataset sources to overcome in future work.
 
\textbf{Model snapshot.}
The 10 evaluated models represent a snapshot of the model landscape as of early 2026.
Absolute vulnerability exposure rates and model rankings may change as models are updated.
To alleviate this, we included the Qwen3 and Qwen3.5 model families in the experiments, and the findings are consistent.
The structural findings, including the temporal lag, the grounding gap, and the ineffectiveness of language-level safety instructions, are more likely to generalize, as they reflect overall characteristics of LLMs instead of a single one.


\section{Related Work}
\label{sec:related}

\subsection{LLM Code Quality and Package Selection}
\label{sec:related_quality}

LLM-generated code introduces risks successively: whether the generated \emph{logic} is secure, whether the \emph{packages} it imports exist, and whether the \emph{versions} of those packages are safe and installable.
Prior work has addressed the first two levels extensively, and this paper addresses the third.

\textbf{Logic-level security.}
Early studies established that GitHub Copilot produces snippets containing CWE-classified weaknesses at non-trivial rates, spanning SQL injection, hard-coded credentials, and cryptographic misuse~\cite{fu2025security}.
Subsequent large-scale analyses confirmed that a significant fraction of generated snippets trigger CWE-classified weaknesses~\cite{dai2025comprehensive,zhao2025cwe,gao2025survey}, and Sajadi \textit{et al.}\ showed that security guidance is frequently absent from LLM responses to Stack Overflow-style questions~\cite{sajadi2025security}.
On the mitigation side, proposed interventions include instruction tuning on security-aware datasets~\cite{he2024instruction,hasan2025teaching}, constrained decoding guided by static analysis~\cite{li2024cosec,fu2024constrained,li2025codistill}, and RAG-augmented prompting with security guidelines~\cite{lin2025rag}.
Wang \textit{et al.}\ characterized security risks in LLM-generated code found in public GitHub repositories~\cite{wang2025aicode}.
None of these mitigations addresses the version dimension: even logically secure code can expose its runtime environment to catalogued vulnerabilities through the dependency versions it pins.

\textbf{Package existence and recommendation.}
Package confusion is a known software supply chain threat~\cite{10.5555/3620237.3620430}, and LLM interfaces have made it more consequential~\cite{10.1145/3728894}.
Spracklen \textit{et al.}\ conducted the most comprehensive study of package hallucination, evaluating 16 LLMs on 576,000 generated samples and finding hallucinated package names at rates from 5.2\% for commercial models to 21.7\% for open-source models~\cite{spracklen2025package}.
Latendresse \textit{et al.}\ examined ChatGPT's library recommendations and showed that models frequently recommend outdated or contextually inappropriate packages~\cite{latendresse2024chatgpt,latendresse2025robust}.
Sipio \textit{et al.}\ documented a popularity bias in LLM-based TPL recommendations~\cite{sipio2025popularity}.
These studies establish that LLM package selection is systematically biased and occasionally hallucinatory, but none examines \emph{version} specifications.
A correctly named, non-hallucinated package can still expose users to risk if the specified version carries a known CVE.
Our work fills this gap: assuming the package name is correct, do LLMs choose a version that is safe and installable?

\subsection{Version Awareness in LLM-Assisted Development}
\label{sec:related_version}

\begin{table}[t]
\centering
\caption{Comparison of related studies along key dimensions.
  \ding{51}: the study addresses this dimension,
  \ding{55}: the dimension is outside the study's scope.}
\label{tab:related_comparison}
\resizebox{\linewidth}{!}{
\setlength{\tabcolsep}{5pt}
\begin{tabular}{lcccccc}
\toprule
\textbf{Dimension} &
  \textbf{Fu \textit{et al.}~\cite{fu2025security}} &
  \textbf{Spracklen \textit{et al.}~\cite{spracklen2025package}} &
  \textbf{Latendresse \textit{et al.}~\cite{latendresse2024chatgpt}} &
  \textbf{Wang \textit{et al.}~\cite{wang2025library}} &
  \textbf{Wu \textit{et al.}~\cite{wu2024versicode}} &
  \textbf{This work} \\
\midrule
Logic security         & \ding{51} & \ding{55} & \ding{55} & \ding{55} & \ding{55} & \ding{55} \\
Package existence      & \ding{55} & \ding{51} & \ding{51} & \ding{55} & \ding{55} & \ding{51} \\
Version safety         & \ding{55} & \ding{55} & \ding{55} & \ding{55} & \ding{55} & \ding{51} \\
Version compatibility  & \ding{55} & \ding{55} & \ding{55} & \ding{51} & \ding{51} & \ding{51} \\
Spontaneous selection   & \ding{55} & \ding{51} & \ding{51} & \ding{55} & \ding{55} & \ding{51} \\
\bottomrule
\end{tabular}
}
\end{table}

A separate line of work examines whether LLMs can correctly reason about specific library versions.
Two research directions have emerged: one studies whether generated code is compatible with a \emph{given} version, the other studies whether generated code can be adapted when APIs evolve.
Both directions presuppose that the version is externally provided or already known.
Neither asks whether the LLM's \emph{spontaneous} version choice is safe and installable.

\textbf{Compatibility with a given version.}
Wang \textit{et al.}\ evaluated LLM-based code completion on tasks involving deprecated APIs, finding that models recommend deprecated usage even when safer alternatives exist in more recent versions, with training data recency as the primary driver~\cite{wang2025library}.
Wu \textit{et al.}\ introduced VersiCode, a benchmark spanning 300 libraries and 2,207 versions, and showed that even GPT-4o scores more than 50 points lower than on standard benchmarks when the task requires version-specific code completion~\cite{wu2024versicode}.
Islah \textit{et al.}\ introduced GitChameleon, an execution-based benchmark of 116 Python problems each conditioned on a specific library version, finding GPT-4o achieves only 39.9\% pass@10 on version-correct generation~\cite{islah2024gitchameleon}.
Misra \textit{et al.}\ extended this with GitChameleon 2.0 that has 328 problems with executable unit tests, finding that state-of-the-art systems achieve only 48--51\% success rates on version-conditioned generation~\cite{misra2025gitchameleon}.
Kuhar \textit{et al.}\ proposed LibEvolutionEval and demonstrated that model accuracy in version-specific generation degrades for recently released versions with limited training coverage~\cite{kuhar2025libevolution,kuhar-etal-2025-libevolutioneval}.

\textbf{Adaptation to API evolution.}
Wu \textit{et al.}\ showed that environment-unaware code generation produces non-executable code at high rates when the deployment environment diverges from training-time assumptions~\cite{wu2026environment}.
Liu \textit{et al.}\ constructed CodeUpdateArena to benchmark knowledge editing for API changes, evaluating how well models adapt code when an API evolves between versions~\cite{liu2025codeupdatearena}.
Work on compatibility failure analysis at the ecosystem level~\cite{11029774,11224651,11121691,10.1145/3808230} further documents the scale of API-breaking changes across library releases.

\textbf{Positioning.}
These benchmarks address a logically downstream question: \emph{given a specified version, can the LLM generate compatible code?}
Our work addresses the logically prior question: \emph{do LLMs choose secure and compatible versions in the first place?}
Version selection precedes code generation, and both must be answered to fully characterize the risk profile of LLM-assisted dependency management.
Unlike benchmark-centric evaluations that present models with explicit version constraints on curated tasks, our study measures spontaneous version-annotation behavior on \myso, directly capturing what developers receive from LLMs under normal usage conditions.
Table~\ref{tab:related_comparison} summarizes how our work differs from the most closely related studies.

 
\section{Conclusion}
\label{sec:conclusion}
 
LLMs are now a primary interface through which developers specify software dependencies, yet the security and compatibility of the version choices they produce have received no systematic study.
We evaluated ten LLMs on \myso and identified three patterns that together characterize version selection as a previously overlooked risk surface.

First, version-annotation behavior is governed by format affordance rather than consistent engineering intent.
Inline prompting elicits specification rates up to 95.18\%, while explicit mode produces as few as 6.45\%.
Across both modes, version choices concentrate on a narrow band of popular releases that lags each model's knowledge cutoff by 9 to 31 months.

Second, the specified versions are systematically vulnerable.
Under inline prompting, 36.70\%--55.70\% of tasks carry at least one CVE, with 62.75\%--74.51\% of vulnerable versions rated \textit{Critical} or \textit{High} severity.
The same small set of vulnerable TPL and version pairs appears across all ten model families, driven by a shared training-signal bias rather than individual model error.

Third, the specified versions are frequently incompatible.
Static compatibility rates fall to 19.70\%--63.20\% under inline prompting, and dynamic BigCodeBench pass rates collapse to 6.49\%--48.62\%.
A controlled neighboring-version experiment confirms that 17\%--55\% of installation failures are attributable to version selection rather than to code-logic defects.

In the near term, practitioners should treat LLM-specified versions as unverified suggestions and route them through validation before committing to a dependency manifest.
The broader implication is that the field must extend its trust model for LLM-generated code beyond logical correctness to encompass the supply-chain dimensions that dependency versions reach into.
We disclosed these findings to all evaluated model providers and to major AI coding assistant platforms. Several parties confirmed the issue and passed the information to the tech team for further improvements.
As AI-assisted development increasingly generates dependency annotations at scale, version selection will account for a growing share of the attack surface that developers inadvertently accept.



\bibliographystyle{ACM-Reference-Format}
\bibliography{llm_tpl}
\end{document}